\documentclass[journal]{IEEEtran}

\usepackage{graphicx}
\usepackage{amsmath,amssymb}
\usepackage{cite}
\usepackage{url}
\usepackage{booktabs}
\usepackage{multirow}
\usepackage{subfig}
\usepackage{tikz}
\usetikzlibrary{positioning}
\usepackage{soul}
\usepackage[hidelinks]{hyperref}
\begin{document}

\title{VolTune: A Fine-Grained Runtime Voltage Control Architecture for FPGA Systems
\thanks{This work is partially supported by JSPS KAKENHI Grant Number ZBA2029720A.}
}

\author{\IEEEauthorblockN{ Akram BEN AHMED, Takahiro HIROFUCHI, Takaaki FUKAI\\}
\IEEEauthorblockA{\textit{Intelligent Platforms Research Institute} \\
\textit{National Institute of Advanced Industrial Sciences and Technology}\\
2-3-26 Aomi, Koto-ku, Tokyo, JAPAN\\
\{akram.benahmed,t.hirofuchi,takaaki.fukai\}@aist.go.jp}
}

\maketitle

\begin{abstract}
The rapid emergence of edge computing platforms and large-scale data centers has made power efficiency a primary design constraint, particularly for data-intensive and AI-driven workloads. Field-programmable gate arrays (FPGAs) are increasingly adopted due to their flexibility and potential for energy-efficient acceleration. However, FPGA supply voltages are typically fixed at design time using conservative margins, limiting the ability to adapt power consumption to runtime conditions.

This paper presents VolTune, an open-source runtime voltage control architecture that enables runtime tuning of FPGA supply voltages through FPGA-integrated control logic that abstracts low-level PMBus operations. VolTune provides both hardware-based and software-based control paths, allowing designers to balance deterministic low-latency operation against programmability. 

In the presented prototype, the hardware-based control path achieves a measured end-to-end voltage transition latency of 2.3 ms, while the controller adds under 2\% static power overhead and under 2\% FPGA resource overhead. 
As a representative case study, VolTune is evaluated on the GTX transceiver supply rail of a Kintex-7 platform. The results show that runtime voltage tuning exposes a bounded operating region with clear trade-offs between energy efficiency and reliability, and achieves up to approximately 29.3\% rail-power reduction at 10.0 Gbps when allowing BER up to $10^{-6}$. These results show that FPGA-integrated runtime voltage control can provide practical energy savings with low integration overhead. 
\end{abstract}



\section{Introduction}

The rapid growth of edge computing platforms and large-scale data centers has placed power efficiency at the center of system design \cite{shi2016edge, satyanarayanan2017emergence}.
Emerging applications such as data analytics, machine learning, and real-time signal processing demand high performance under strict energy and thermal constraints, particularly in distributed edge-cloud environments where power and cooling budgets are inherently limited \cite{satyanarayanan2017emergence}.
As a result, architectural flexibility and energy-efficient execution have become critical requirements across a wide range of computing environments.

Field-programmable gate arrays (FPGAs) have gained significant traction in these domains due to their reconfigurability, high throughput, and favorable performance-per-watt characteristics compared to general-purpose processors \cite{mittal2020cnn_survey}.
Major technology companies have made sustained investments in FPGA-based infrastructure for both data center and edge deployments, integrating FPGAs into production systems for acceleration, networking, and system offload, and positioning them as a key building block for future edge-cloud architectures \cite{putnam2014reconfigurable,firestone2018azure,aws2023nitro_whitepaper}.

Despite this growing role, power management in FPGA-based systems has not evolved at the same pace as architectural capabilities.
In current practice, FPGA supply voltages are typically fixed at design time using conservative margins to guarantee correct operation across worst-case process, voltage, and temperature conditions.
Commercial FPGA platforms such as Xilinx and Intel devices specify nominal core and transceiver voltages with tight operating ranges, leaving limited opportunity for runtime adaptation within standard design flows \cite{xilinx_ultrascale_ds,intel_stratix10_ds}. While this approach simplifies validation, it prevents systems from adapting power consumption to runtime conditions, workload variation, or application-level tolerance.

In practice, workload characteristics in edge and data center systems vary significantly over time, including changes in computation intensity, memory access patterns, and communication activity \cite{barroso2013datacenter}.
During low-utilization or communication-light phases, operating all FPGA voltage rails at worst-case margins results in unnecessary dynamic and static power consumption.
Prior studies have demonstrated that supply voltage reduction can yield quadratic reductions in dynamic power, with measured FPGA power savings exceeding 20--40\% under controlled undervolting scenarios \cite{salami2018micro_undervolting}.
Moreover, many applications, such as machine learning inference, multimedia processing, and approximate data analytics, exhibit inherent tolerance to bounded errors or minor performance variations, enabling operation at reduced voltage without catastrophic failure \cite{han2013approximate,mittal2016survey}.
Flexible voltage control provides an architectural mechanism that can be leveraged to exploit temporal slack and application-level tolerance in pursuit of improved energy efficiency.

Although modern FPGA platforms provide programmable voltage regulators accessible through standardized interfaces such as PMBus \cite{pmbus}, these interfaces are typically exposed at the board-management or software level and require low-level command sequences.
The resulting control latency and integration complexity make frequent or fine-grained runtime voltage adjustments difficult to integrate within standard FPGA design flows.
Consequently, runtime voltage tuning remains rarely integrated as a first-class architectural mechanism in FPGA-based systems.

This paper addresses this gap by proposing \emph{VolTune}, a runtime voltage control architecture that integrates voltage control directly into FPGA logic. 
VolTune abstracts low-level PMBus operations behind a structured control interface, enabling voltage adjustment to be triggered at runtime through hardware or software control paths.
Unlike prior undervolting approaches, VolTune is explicitly designed for runtime use, supports fine-grained voltage adjustment within regulator resolution limits, and characterizes control overhead and transition behavior.

The contributions of this work are summarized as follows:
\begin{itemize}
    \item An FPGA-integrated voltage-control architecture for runtime supply tuning, including a structured abstraction of PMBus-based regulation for hardware and software control paths.
    \item A detailed characterization of the VolTune controller, including control latency, voltage transition behavior, and hardware overhead.
    \item An experimental evaluation of runtime voltage tuning on the transceiver supply rail of a Kintex-7 platform.
    \item To support reproducibility and follow-on use, the VolTune implementation is released as an open-source repository containing the design artifacts used in this work \cite{voltune_repo}.
\end{itemize}

The remainder of this paper is organized as follows.
Section~\ref{sec:related} reviews the background and related work.
Section~\ref{sec:architecture} presents the VolTune architecture and system model.
Section~\ref{sec:pmbus_impl} describes the PMBus control engine implementation and associated design trade-offs.
Section~\ref{sec:controller_eval} evaluates the VolTune controller in terms of latency and overhead.
Section~\ref{sec:case_study} presents a representative case study using serial transceivers.
Section~\ref{sec:discussion} discusses limitations and generality.
Finally, Section~\ref{sec:conclusion} concludes the paper.

\section{Technical Background and Related Work}
\label{sec:related}

\subsection{Power Management in FPGA-Based Systems}

Power management in FPGA-based systems has historically relied on static, board-level configuration, rather than closed-loop, runtime adaptation. A central difficulty is that safe operating points depend on the user design’s timing, as well as temperature and voltage droop, which are not known with sufficient precision at design time.

\textit{Chow et al.} \cite{chow2005_dvs_fpt} presented one of the early demonstrations that dynamic voltage scaling can be applied to commercial FPGAs without modifying the silicon, by using an on-chip delay sensor (LDMC) and a closed-loop controller that adjusts the FPGA supply voltage to track timing margin across temperature changes. Their setup uses external voltage control, and the key contribution is the runtime feedback mechanism that links measured delay to voltage selection, rather than a fixed guardband.

\textit{Levine et al.} \cite{levine2014_dvfs_slack_fpga} moved closer to practical DVFS for user designs by proposing online slack measurement insertion (SMI) and using the measured slack to drive dynamic voltage and frequency scaling decisions. This line of work explicitly targets timing monitoring during operation, and highlights that DVFS control is constrained by what can be instrumented in the fabric, with overheads in logic and clocking resources.

\textit{Farhadi et al.} \cite{beldachi2014_zynq_power_control_fpl} demonstrated that implementing PMBus access in programmable logic reduces latency and overhead compared to software execution on the processing system. Their work focuses on monitoring accuracy and control-path comparison. In contrast, VolTune formalizes an FPGA-integrated runtime voltage-control architecture that exposes a structured command abstraction and supports systematic controller characterization and workload-level evaluation.

From a power-delivery and calibration perspective, \textit{Zhao et al.} \cite{zhao2016_universal_dvfs_apec} proposed a self-calibrating DVFS scheme that characterizes voltage and frequency operating limits across temperature points, stores them in a table, and then applies DVFS at runtime according to measured temperature. A key practical aspect is that their method explicitly interacts with a digitally controlled DC-DC converter, and treats converter response and quantization as part of the DVFS design, rather than assuming idealized voltage transitions.

\subsection{Undervolting and Timing Speculation}

Reducing supply voltage below nominal levels is a well-established technique for lowering dynamic and static power consumption, but it introduces timing violations as circuit delays increase. To mitigate this, architectural mechanisms for detecting and recovering from timing errors have been proposed.

\textit{Ernst et al.} \cite{ernst2003razor} introduced Razor, a circuit-level timing speculation mechanism that augments flip-flops with shadow latches to detect late-arriving signals. When a timing error is detected, the pipeline recovers using architectural replay. This work demonstrated that systems can operate beyond conservative voltage margins by dynamically detecting and correcting timing failures, thereby enabling energy savings without permanent functional errors.

Extending this idea, \textit{Das et al.} \cite{das2008_rerazor} proposed enhanced in-situ error detection techniques that improve robustness under aggressive voltage scaling. Their work further established that error-aware voltage scaling can be made practical through microarchitectural support.

In the FPGA domain, undervolting behavior has been studied experimentally rather than through architectural error recovery.
\textit{Salami et al.} \cite{salami2018micro_undervolting} performed a comprehensive evaluation of supply voltage underscaling in FPGA on-chip memories. They characterized failure modes, error rates, and the relationship between voltage reduction and memory reliability, demonstrating measurable power savings at reduced voltage levels while quantifying the onset of errors. In follow-up work, the same authors evaluated the built-in Error-Correction Code (ECC) in FPGA on-chip memories under aggressive undervolting \cite{undervolting2}, showing that the SECDED ECC mechanism can correct over 90\% of undervolting faults in BRAMs and enable further power savings with minimal application accuracy loss when used with neural network accelerators. They later extended their study to reduced-voltage operation across multiple on-chip FPGA components in the context of CNN acceleration \cite{undervolting4}, highlighting significant energy-efficiency gains and characterizing the corresponding reliability behavior across different blocks within the FPGA fabric.

\textit{Ben Ahmed et al.} \cite{Akram_undervolting_2022} presented a preliminary exploration of undervolting for error-permissive serial interconnection in multi-FPGA systems, using KC705 transceivers as a case study. That work examined the relationship between transceiver supply voltage, bit error rate, and power reduction under approximate communication settings, and identified runtime voltage adaptation as a promising direction. However, it did not implement an FPGA-integrated voltage-control architecture, did not abstract PMBus access into a reusable control interface, and did not characterize the latency and overhead of the control path itself. VolTune extends that direction by introducing a concrete runtime voltage-control mechanism that can be invoked from FPGA logic and evaluated independently from the application case study.

Collectively, these empirical studies establish that undervolting can yield significant energy savings in FPGAs, and that fault behavior under aggressive undervolting is characterizable and, in some cases, mitigable. However, these works primarily focus on characterization and mitigation techniques, rather than providing an architectural abstraction that enables systematic runtime voltage control within FPGA logic.

\subsection{Approximate Computing and Error Tolerance}
Approximate computing is based on the observation that many applications can tolerate bounded inaccuracies in exchange for improved energy efficiency. Although many FPGA studies focus on arithmetic datapaths and accelerator kernels, the same broader principle is relevant to communication and system-level behavior when bounded errors can be tolerated without unacceptable degradation in application outcomes.

\textit{Han and Orshansky} \cite{han2013approximate} introduced the concept as a cross-layer design paradigm that relaxes strict correctness to reduce energy. \textit{Mittal} \cite{mittal2016survey} surveyed architectural and circuit-level techniques that trade accuracy for improvements in power, delay, or area, and organized the design space of approximation methods.

These ideas are particularly relevant to FPGA accelerators, because FPGAs allow designers to customize arithmetic datapaths and numeric representations in hardware. \textit{Ullah et al.} \cite{ullah2018_dac_approx_mult} proposed FPGA-oriented approximate multipliers that exploit LUT structures and carry chains, and reported gains in area, latency, and energy relative to vendor multiplier IPs, while bounding accuracy loss. \textit{Perri et al.} \cite{perri2020_fpga_approx_add} designed approximate adders specialized for FPGA LUT resources by approximating the least significant bits, and evaluated error metrics and implementation costs on an Artix-7 device.

Approximation has also been applied at the accelerator level for machine learning. \textit{Korol et al.} \cite{korol2022_confax} proposed a configurable framework that swaps accurate and approximate multipliers in FPGA CNN accelerators to navigate the accuracy-performance-resource trade-off at the edge.

Overall, these FPGA-focused studies support the practical premise that controlled approximation can be implemented in FPGA fabrics at the arithmetic and accelerator levels. More broadly, they support the view that bounded-error operation can be a useful design point when application-level behavior remains acceptable. This motivates runtime mechanisms that can exploit application-level tolerance while keeping system behavior within acceptable bounds.

\subsection{Positioning and Architectural Gap}

While prior work has demonstrated dynamic voltage scaling using delay monitoring, architectural timing speculation, and empirical undervolting characterization, these approaches do not provide a reusable FPGA-integrated control abstraction for structured runtime rail tuning. Similarly, prior FPGA approximation studies mainly focus on arithmetic units and accelerator-level design choices, rather than runtime control of board-level supply rails for communication-oriented operating points. What remains insufficiently explored is a reusable architectural abstraction that integrates board-level voltage regulation directly into FPGA logic in a structured and runtime-invocable manner. VolTune addresses this gap by embedding a structured voltage-control interface within the FPGA platform, abstracting low-level PMBus transactions and enabling systematic controller characterization and workload-level evaluation under dynamic voltage conditions.

\section{VolTune Architecture}
\label{sec:architecture}

\subsection{Design Objectives and Scope}
\label{sec:arch_objectives}

VolTune is designed to expose board-level voltage control as a runtime architectural mechanism within FPGA systems. Rather than relying on external firmware or manual configuration, it integrates voltage-control functionality into the FPGA fabric through a structured interface that can be coordinated with application execution and system state. The design is guided by the following principles:

\begin{itemize}
\item \textbf{Runtime Invocation:}
Voltage transitions must be triggerable during system operation, without requiring reconfiguration of the FPGA bitstream or manual board-level intervention. The architecture therefore provides a command-driven interface that can be invoked programmatically at runtime through either hardware or software control paths.
\item \textbf{Rail-Level Granularity:}
VolTune operates at the level of discrete supply rails exposed by the board’s programmable regulator.
It does not modify the internal silicon power grid or attempt fine-grained cell-level control.
Voltage adjustments are constrained to the resolution and limits supported by the external regulator.
\item \textbf{FPGA-Resident Control Path:}
VolTune places the voltage-control functionality inside the FPGA platform rather than relying on an external host-side control loop. Depending on the implementation, command sequencing can be realized either directly in FPGA logic or through an embedded processor subsystem, allowing different trade-offs between determinism and flexibility.
\item \textbf{Abstraction of Low-Level Transactions:}
Board-level voltage regulators are commonly accessed through PMBus command sequences. VolTune abstracts these low-level protocol transactions behind a structured control interface, allowing higher-level logic or software to request voltage changes without directly managing PMBus signaling details.
\item \textbf{Explicit Characterization of Transition Behavior:}
Voltage scaling is treated as a measurable and characterizable process.
The architecture accounts for command transmission latency, regulator response time, and settling behavior, rather than assuming instantaneous voltage changes.
\end{itemize}

The scope of VolTune is intentionally limited.
It does not implement circuit-level timing speculation or error-recovery mechanisms.
It does not modify the FPGA silicon or internal voltage distribution network.
It does not assume application-level fault tolerance.
Instead, VolTune provides a reusable architectural layer that exposes controlled, runtime voltage adjustment within the operational limits specified by the FPGA vendor and regulator hardware.
This scoped design allows VolTune to remain portable across platforms that provide digitally programmable regulators, while preserving compatibility with standard FPGA tool flows.

Within VolTune, the core voltage-control functionality is implemented as a PowerManager subsystem, with either a hardware or software realization. Both realizations expose a common control model, while their implementation-specific trade-offs are discussed in the following subsections.

\begin{figure}[t]
\centering
\includegraphics[width=\linewidth]{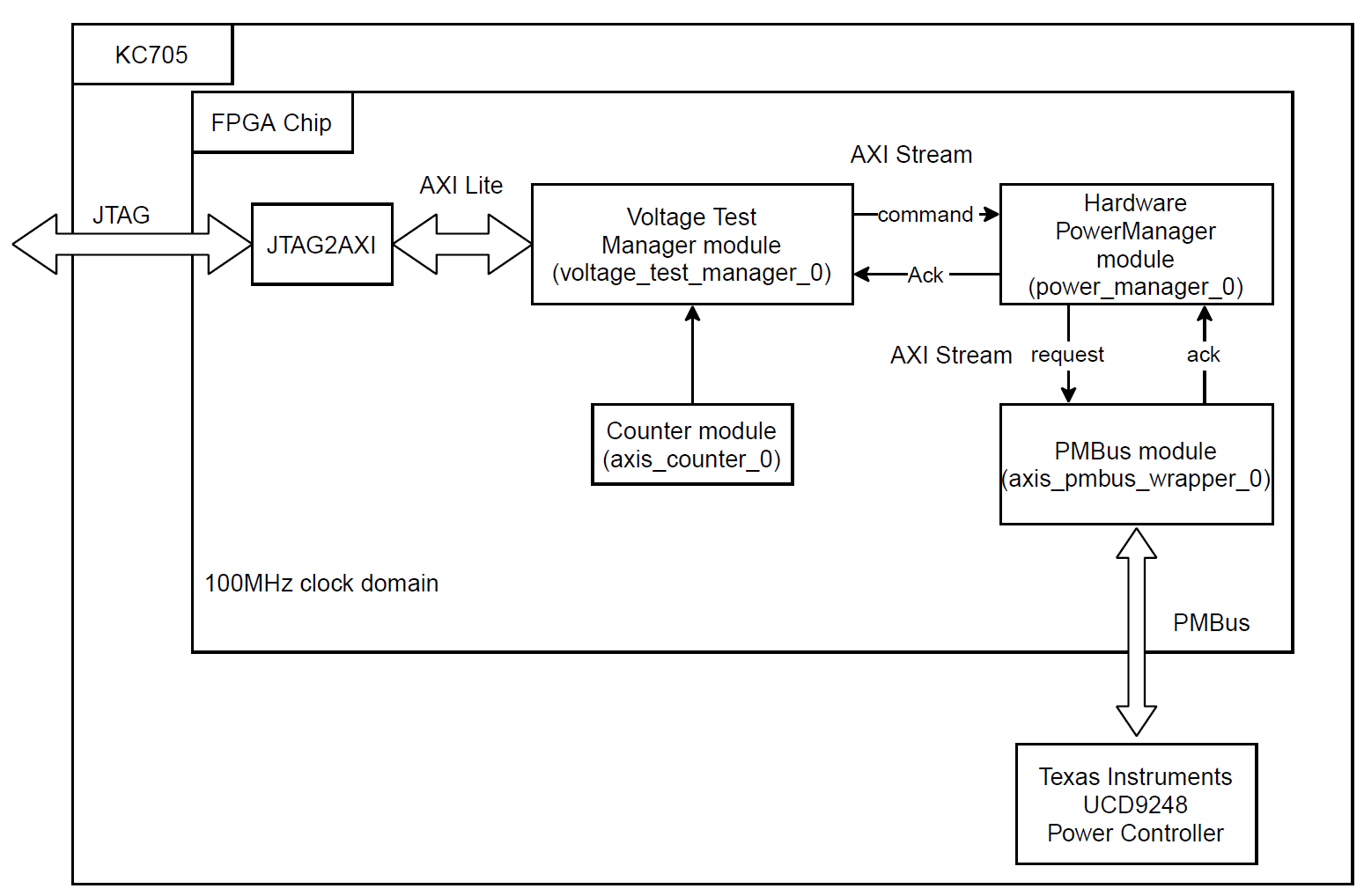}
\caption{VolTune hardware implementation block diagram}
\label{fig:powermanager_arch}
\end{figure}

\subsection{VolTune Hardware Implementation Overview}
\label{sec:arch_hw_overview}

Figure~\ref{fig:powermanager_arch} shows the block diagram of the hardware realization of VolTune on the KC705 platform. The realization includes the core VolTune control path together with auxiliary modules used for configuration and evaluation in the prototype implementation.

\subsubsection*{\textbf{Voltage Test Manager Module}}
The Voltage Test Manager module represents the application side of the system and can host the main FPGA workload, such as AI inference or transmission-related processing. It interfaces with the PowerManager subsystem whenever runtime voltage adjustments are required.

In the prototype implementation shown in Figure~\ref{fig:powermanager_arch}, the same module is also used to orchestrate controlled voltage-transition experiments by issuing voltage-change commands to the Hardware PowerManager through an AXI Stream interface. For configuration and evaluation, external register access is provided through the JTAG2AXI path shown in the figure, while a Counter module in the same clock domain is used to support repeatable timing measurements during latency characterization.

\subsubsection*{\textbf{Hardware PowerManager Module}}
The Hardware PowerManager module performs the main control operations inside the FPGA fabric. It receives command messages from the Voltage Test Manager module through the AXI Stream interface. These commands request actions such as voltage updates and telemetry readback.

Upon receiving a command, the Hardware PowerManager module translates the request into a sequence of operations directed to the PMBus module. It also receives acknowledgment signals from the PMBus module and forwards the corresponding status information back to the Voltage Test Manager module.

\subsubsection*{\textbf{PMBus Module}}
The PMBus module serves as the low-level transaction engine between the Hardware PowerManager and the external programmable power controller. It executes PMBus transactions over the SCL and SDA lines and returns acknowledgment or readback information to the Hardware PowerManager. In the prototype platform, this interface connects to a Texas Instruments UCD9248 multi-rail programmable power controller \cite{ti_ucd9248}, which applies the requested voltage updates to the selected board-level supply rail. The internal interface and transaction behavior are detailed in Section~\ref{sec:pmbus_impl}.

\begin{figure*}[t]
\centering
\includegraphics[width=\linewidth]{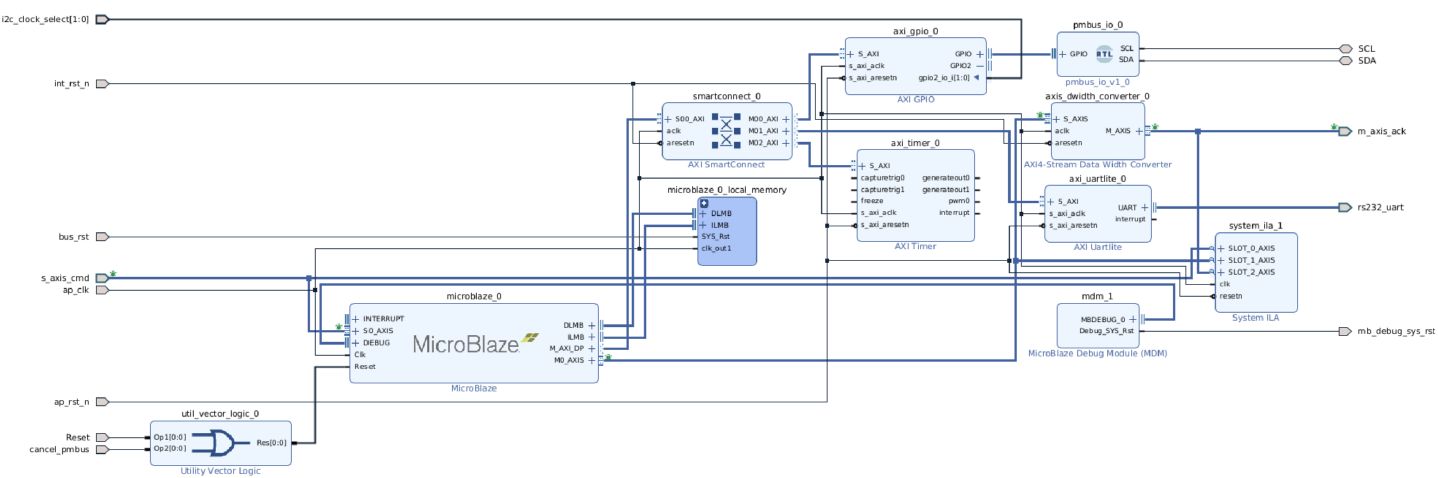}
\caption{Software implementation block design snapshot from Vivado}
\label{fig:sw_powermanager_bd}
\end{figure*}
\begin{figure}[t]
\centering
\includegraphics[width=\linewidth]{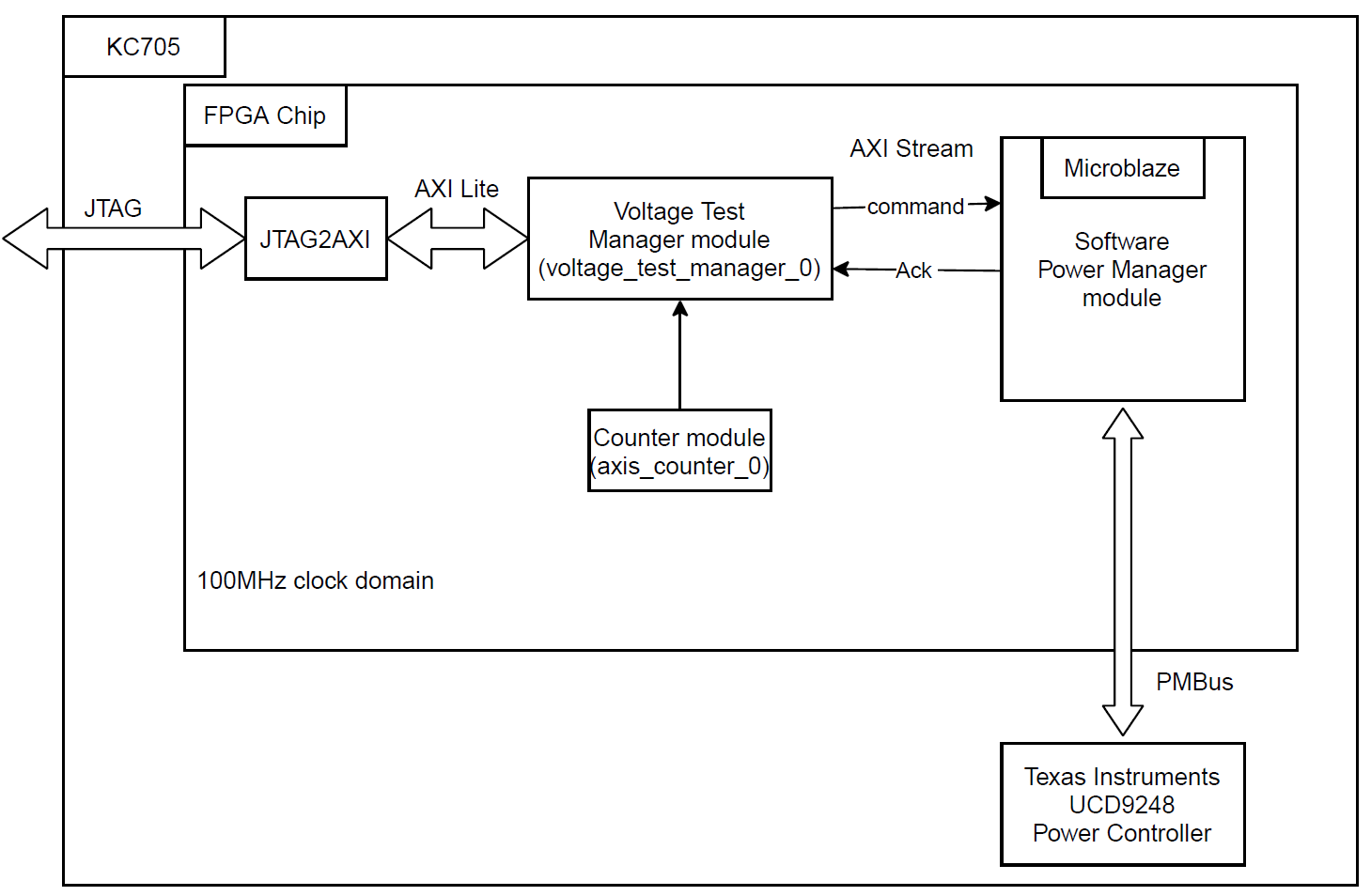}
\caption{VolTune software implementation block diagram}
\label{fig:sw_powermanager}
\end{figure}

\subsection{VolTune Software Implementation Overview}
\label{sec:arch_sw_overview}

VolTune also includes a software realization of the PowerManager subsystem, referred to as the Software PowerManager. In this design, the PowerManager logic executes on a MicroBlaze soft processor while maintaining the same high-level command semantics as the hardware implementation.

Figure~\ref{fig:sw_powermanager_bd} shows the complete software implementation architecture as captured from the AMD Xilinx Vivado design environment, while Figure~\ref{fig:sw_powermanager} provides a simplified block-diagram view of the same organization. In this realization, a MicroBlaze-based control subsystem issues voltage-control requests in software and drives the same PMBus-controlled regulator used in the hardware implementation. These requests are translated into PMBus transactions issued to the external Texas Instruments UCD9248 Power Controller \cite{ti_ucd9248} over the PMBus interface \cite{pmbus}.

As shown in Figure~\ref{fig:sw_powermanager_bd}, the MicroBlaze processor is connected to memory and peripheral modules through an AXI-based interconnect. PMBus access is realized through AXI-connected peripherals, enabling the software to issue the required transaction sequences and receive acknowledgment status.

\subsection{Design Trade-offs}
\label{sec:arch_tradeoffs}

The hardware realization favors deterministic sequencing and a more direct control path, which is useful when voltage control must be tightly coupled with runtime FPGA activity. The software realization favors flexibility and ease of modification, but requires additional infrastructure, including a processor subsystem, memory, and AXI-connected peripherals, which increases resource usage and can add software and subsystem latency to the control path. In both cases, overall voltage-transition behavior remains constrained by serialized PMBus transactions and external regulator settling time, which motivates the controller characterization presented in the next section.

\section{PMBus Control Engine Implementation}
\label{sec:pmbus_impl}

This section describes how VolTune controls KC705 voltage rails through the TI UCD9248 power controller \cite{ti_ucd9248} using PMBus \cite{pmbus}. As introduced in Section~\ref{sec:architecture}, the PMBus module serves as the low-level transaction engine between the Hardware PowerManager and the external power controller. Here, we focus on the transaction primitives, the request conversion path, and the concrete command sequences used for voltage programming and readback on KC705.

For simplicity, VolTune implements only the subset of PMBus commands required by the target workflows, rather than the full PMBus command language. In this section, the PMBus control engine and command translation logic are implemented in RTL and HLS, so additional PMBus commands can be added by extending the command translation layer when needed. To avoid confusion between control layers, we refer to the internal command identifiers exchanged between Voltage Test Manager and PowerManager as VolTune opcodes, while the standardized PMBus command bytes transmitted to the UCD9248 are referred to as PMBus commands.

\subsection{PMBus Basics}
\label{sec:pmbus_basics}
\begin{table*}[t]
\centering
\caption{Subset of PMBus commands used by VolTune.}
\label{tab:pmbus_cmd_subset}
\begin{tabular}{c l l l}
\toprule
\textbf{Code} & \textbf{Command name} & \textbf{R/W} & \textbf{Typical use} \\
\midrule
00h & \texttt{PAGE} & R/W & Select target rail (multi-rail devices) \\
03h & \texttt{CLEAR\_FAULTS} & W & Clear latched fault status \\
21h & \texttt{VOUT\_COMMAND} & R/W & Program output voltage setpoint \\
43h & \texttt{VOUT\_UV\_WARN\_LIMIT} & R/W & Set undervoltage warning threshold \\
44h & \texttt{VOUT\_UV\_FAULT\_LIMIT} & R/W & Set undervoltage fault threshold \\
5Eh & \texttt{POWER\_GOOD\_ON} & R/W & Set power-good on threshold \\
5Fh & \texttt{POWER\_GOOD\_OFF} & R/W & Set power-good off threshold \\
8Bh & \texttt{READ\_VOUT} & R & Read back output voltage \\
8Ch & \texttt{READ\_IOUT} & R & Read output current telemetry \\
\bottomrule
\end{tabular}
\end{table*}
\begin{figure}[t]
\centering
\subfloat[Send Byte]{\includegraphics[width=\linewidth]{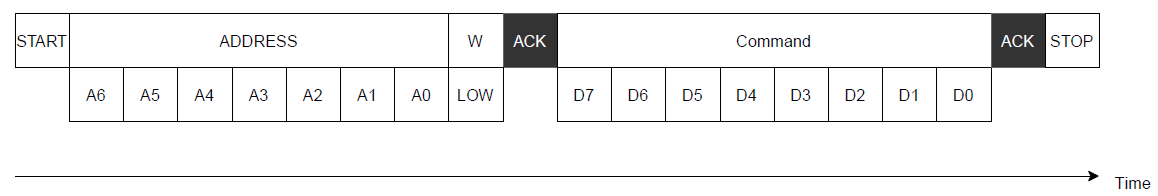}\label{fig:pmbus_send_byte}}\\
\vspace{0.6em}
\subfloat[Write Byte and Write Word]{\includegraphics[width=\linewidth]{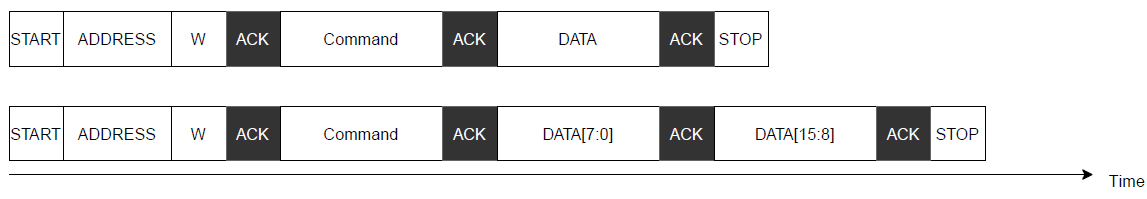}\label{fig:pmbus_write}}\\
\vspace{0.6em}
\subfloat[Read Byte and Read Word]{\includegraphics[width=\linewidth]{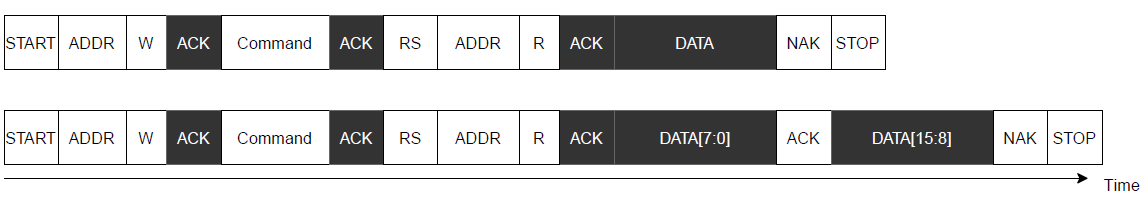}\label{fig:pmbus_read}}
\caption{PMBus transaction primitives.}
\label{fig:pmbus_primitives}
\end{figure}

PMBus is a power-management protocol that uses an I$^2$C-compatible two-wire bus. Communication occurs on SCL and SDA, where each transaction is framed by a START condition and a STOP condition, as summarized in Figure~\ref{fig:pmbus_primitives}. The START condition is defined by SDA transitioning from 1 to 0 while SCL is high, and the STOP condition is defined by SDA transitioning from 0 to 1 while SCL is high. A transaction begins with a 7-bit device address followed by a single R/W bit, and each transferred byte is followed by an acknowledgment bit on the ninth clock pulse \cite{pmbus}.

PMBus operations are identified by an 8-bit PMBus command. For write operations, the master sends the PMBus command followed by one or more data bytes, as shown in Figure~\ref{fig:pmbus_write}. For read operations, the master first writes the PMBus command, then performs a read transaction to receive one or more data bytes \cite{pmbus}, as shown in Figure~\ref{fig:pmbus_read}. PMBus defines a large standardized command set, so Table~\ref{tab:pmbus_cmd_subset} lists the subset of PMBus commands used by VolTune in the current prototype, while the complete command-language definition is provided in the PMBus Part II specification \cite{pmbus_spec_part2_v1_1}.

Many PMBus controllers, including multi-rail devices such as the UCD9248, multiplex multiple outputs behind a shared bus interface. Rail selection is performed using the PAGE mechanism, where the PAGE command sets an internal selector so that subsequent PMBus commands apply to the chosen rail. In practice, the controller issues PAGE when the target rail changes, then applies voltage-programming or telemetry commands for that rail \cite{pmbus}.

\subsection{PMBus Module Interface in VolTune}
\label{sec:pmbus_iface}

As introduced in Section~\ref{sec:architecture}, the PMBus module serves as the low-level transaction engine between the Hardware PowerManager and the external Texas Instruments UCD9248 Power Controller. In VolTune, each PMBus command, together with any associated payload bytes, is packaged into a stream-level request sent from the PowerManager to the PMBus module. The module then generates the corresponding bus-level signaling on SCL and SDA, including START and STOP conditions and byte-level ACK/NACK handshaking. For read operations, the module returns the received data bytes and completion status through its response channel, allowing the PowerManager to interpret the result and propagate it back to the requester. Protocol failures detected during execution are reported through structured status signals, enabling deterministic error handling without exposing bus timing details to higher-level logic.

In the present implementation, the PMBus module is operated at two clock rates, 100\,kHz and 400\,kHz. These PMBus clock settings are important design choices because they directly affect transaction time and therefore influence control latency and measurement granularity. Their impact on controller behavior is evaluated later in Section~\ref{sec:controller_eval}.

PMBus payload values are encoded using fixed-point formats. In this design, voltage programming and readback use \texttt{LINEAR16}, while telemetry may use formats such as \texttt{LINEAR11}, depending on the PMBus command and device behavior \cite{pmbus,ti_ucd9248}. In the present prototype, this voltage encoding is matched to the UCD9248 configuration used on KC705. More generally, the exact encoding of voltage-related PMBus payloads depends on the target regulator configuration, including the interpretation of \texttt{VOUT\_MODE}. Porting VolTune to a different PMBus-controlled platform may therefore require adapting the voltage encoding and decoding path to the target device and board configuration.

\subsection{Rail Selection on KC705}
\label{sec:pmbus_rail_map}

VolTune treats each supply rail as a selectable target identified by a lane number. The lane number is a VolTune-specific identifier used by the PowerManager and is not part of the PMBus standard. For each request, the PowerManager translates the lane number into a PMBus device address and a PMBus \texttt{PAGE} value, which together select the corresponding regulator output channel.

Table~\ref{tab:kc705_lane_map} lists the mapping used in our KC705 implementation. During operation, the controller issues a \texttt{PAGE} update when the target lane changes, then applies subsequent PMBus commands to the selected rail. The same approach can be adopted on other PMBus-controlled multi-rail FPGA platforms by providing the corresponding lane-to-(address,\texttt{PAGE}) mapping for the target board and regulator configuration.
\begin{table}[t]
\centering
\caption{KC705 rail mapping.}
\label{tab:kc705_lane_map}
\begin{tabular}{c l c c}
\toprule
\textbf{Lane} & \textbf{Rail name} & \textbf{PMBus addr.} & \textbf{PAGE} \\
\midrule
0  & VCCINT      & 52 & 0 \\
1  & VCCAUX      & 52 & 1 \\
2  & VCC3V3      & 52 & 2 \\
3  & VADF        & 52 & 3 \\
4  & VCC2V5      & 53 & 0 \\
5  & VCC1V5      & 53 & 1 \\
6  & MGTAVCC     & 53 & 2 \\
7  & MGTAVTT     & 53 & 3 \\
8  & ACCAUX\_IO  & 54 & 0 \\
9  & VCCBRAM     & 54 & 1 \\
10 & MGTVCCAUX   & 54 & 2 \\
\bottomrule
\end{tabular}
\end{table}

\subsection{Command Packing and Conversion Path}
\label{sec:pmbus_packing}

\begin{figure}[t]
\centering
\includegraphics[width=\linewidth]{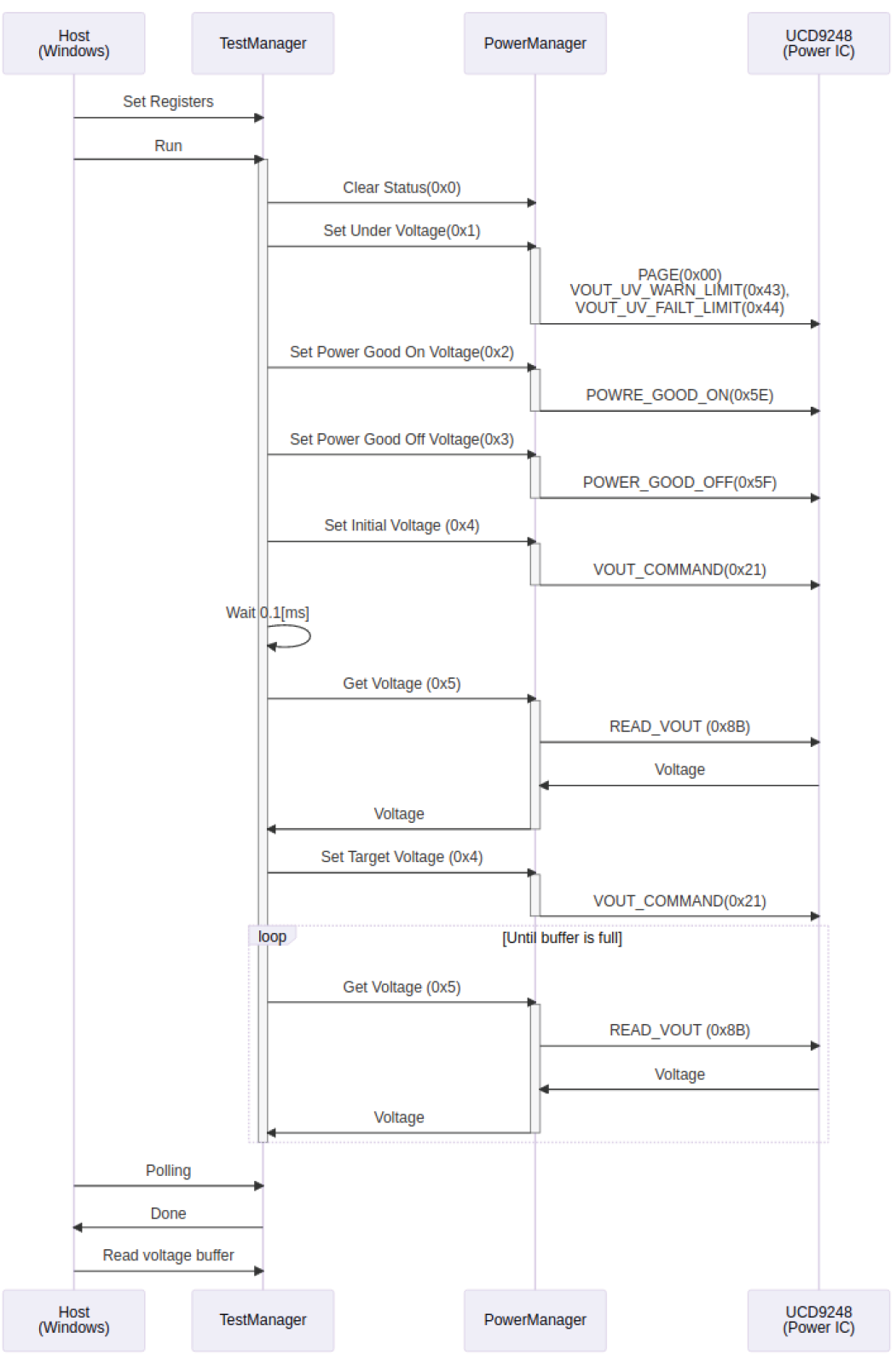}
\caption{Prototype voltage measurement sequence.}
\label{fig:voltage_measure_seq}
\end{figure}

At the application side, VolTune accepts structured requests in the form of VolTune opcodes together with a target lane and an associated value. In the prototype workflow, these requests are generated by the test infrastructure and delivered to the PowerManager through the Voltage Test Manager, as illustrated in Figure~\ref{fig:voltage_measure_seq}. The PowerManager acts as the translation layer between control software and the PMBus bus, converting each VolTune opcode request into one or more PMBus commands issued to the UCD9248.
\begin{table*}[t]
\centering
\caption{VolTune opcode to PMBus command mapping.}
\label{tab:opcode_map}
\begin{tabular}{c l l}
\toprule
\textbf{VolTune opcode} & \textbf{Operation requested from PowerManager} & \textbf{PMBus command(s) issued to UCD9248} \\
\midrule
0x0 & Clear Status      & Internal state clear --- (no PMBus transaction, controller-internal reset)\\
0x1 & Set Under Voltage & \texttt{PAGE} (0x00, when lane changes) \\
    &                   & \texttt{VOUT\_UV\_WARN\_LIMIT} (0x43) \\
    &                   & \texttt{VOUT\_UV\_FAULT\_LIMIT} (0x44) \\
0x2 & Set Power Good On Voltage & \texttt{POWER\_GOOD\_ON} (0x5E) \\
0x3 & Set Power Good Off Voltage & \texttt{POWER\_GOOD\_OFF} (0x5F) \\
0x4 & Set Voltage & \texttt{VOUT\_COMMAND} (0x21) \\
0x5 & Get Voltage & \texttt{READ\_VOUT} (0x8B) \\
\bottomrule
\end{tabular}
\end{table*}
The conversion path follows three steps:
\begin{enumerate}
    \item The PowerManager resolves the target lane into a PMBus device address and a \texttt{PAGE} value using the mapping in Table~\ref{tab:kc705_lane_map}.
    \item It selects the appropriate PMBus transaction type, for example Write Word for voltage and threshold programming or Read Word for voltage and telemetry readback, consistent with the PMBus primitives introduced in Section~\ref{sec:pmbus_basics}.
    \item It packs the selected PMBus command and its payload bytes into the PMBus module request stream. Payload values are encoded using the fixed-point formats described earlier, for example \texttt{LINEAR16} for voltage programming and readback and \texttt{LINEAR11} for some telemetry commands.
\end{enumerate}

Each PMBus transaction is executed atomically. The next transaction is issued only after the PMBus module reports completion through its response channel and status signals. This serialized execution model avoids overlapping bus activity and ensures deterministic command sequencing. As illustrated in Figure~\ref{fig:voltage_measure_seq}, a VolTune opcode request to set a rail voltage expands into an ordered PMBus sequence, including a \texttt{PAGE} update when the lane changes, followed by a Write Word transaction that issues the \texttt{VOUT\_COMMAND} PMBus command with a \texttt{LINEAR16}-encoded voltage payload.

\begin{table*}[t]
\centering
\caption{VolTune operation mapping on KC705.}
\label{tab:voltune_pmbus_map}
\begin{tabular}{p{0.18\linewidth} p{0.48\linewidth} p{0.28\linewidth}}
\toprule
\textbf{Operation} & \textbf{PMBus command sequence (conceptual)} & \textbf{Primitive types} \\
\midrule
Select rail & Write \texttt{PAGE} when lane changes & Write Byte \\
\midrule
Voltage update & Prototype workflow: threshold writes, then \texttt{VOUT\_COMMAND} & Write Byte + Write Word (multiple) \\
\midrule
Voltage readback & \texttt{PAGE} when lane changes, then \texttt{READ\_VOUT} & Write Byte + Read Word \\
\midrule
Telemetry readback & \texttt{PAGE} when lane changes, then periodic telemetry reads (\texttt{READ\_IOUT}) & Write Byte + Read Word (periodic) \\
\midrule
Controller reset & Internal state clear only, no PMBus transaction & --- \\
\bottomrule
\end{tabular}
\end{table*}

\subsection{Voltage Update and Readback Sequences}
\label{sec:pmbus_sequences}

This subsection describes the concrete PMBus sequences used for voltage programming and readback. These sequences are implemented using the primitives in Figure~\ref{fig:pmbus_primitives} and the lane-to-(address,\texttt{PAGE}) mapping introduced in Section~\ref{sec:pmbus_rail_map}.

\subsubsection*{\textbf{Voltage update sequence}}

In VolTune, voltage programming is realized as an explicit PMBus sequence rather than a single abstract control action. In the prototype measurement workflow shown in Figure~\ref{fig:voltage_measure_seq}, this sequence includes threshold-register configuration followed by the final voltage-setpoint update. This runtime sequence assumes that the target rail is already enabled and operating in its nominal board configuration before VolTune issues voltage updates. The present work focuses on runtime voltage adjustment after platform bring-up, rather than on a generic PMBus initialization or power-state management sequence.
As a concrete example, consider a request to set the voltage of \texttt{VCCBRAM} to 0.9\,V. From Table~\ref{tab:kc705_lane_map}, \texttt{VCCBRAM} corresponds to lane 9, which maps to PMBus address 54 and \texttt{PAGE}=1. Using the PMBus primitives in Figure~\ref{fig:pmbus_primitives} and the opcode-to-command mapping in Table~\ref{tab:opcode_map}, the resulting PMBus sequence is:

\begin{enumerate}
\item Write Byte: \\
\texttt{[Addr=54][PAGE (00h)][01h]}

\item Write Word: \\
\texttt{[Addr=54][VOUT\_UV\_WARN\_LIMIT (43h)][Threshold]}

\item Write Word: \\
\texttt{[Addr=54][VOUT\_UV\_FAULT\_LIMIT (44h)][Threshold]}

\item Write Word: \\
\texttt{[Addr=54][POWER\_GOOD\_ON (5Eh)][Threshold]}

\item Write Word: \\
\texttt{[Addr=54][POWER\_GOOD\_OFF (5Fh)][Threshold]}

\item Write Word: \\
\texttt{[Addr=54][VOUT\_COMMAND (21h)][LINEAR16(0.9V)]}
\end{enumerate}

In this example, the PMBus address selects the regulator device and the PAGE command selects the target rail within that device. The payloads denoted by \texttt{[Threshold]} represent PMBus-encoded voltage threshold values written to the under-voltage warning, under-voltage fault, and power-good registers. In the prototype workflow, these threshold values are explicitly programmed before the final \texttt{VOUT\_COMMAND} update so that the regulator state remains consistent with the requested rail setting. Their exact values are chosen according to the intended operating point and the corresponding protection or monitoring limits of the target rail. This sequence reflects the prototype measurement workflow, while protocol handling remains internal to the PMBus module.

\subsubsection*{\textbf{Voltage readback sequence}}
Voltage readback is performed using a Read Word transaction with \texttt{READ\_VOUT}, consistent with Figure~\ref{fig:pmbus_read}. The PowerManager waits for transaction completion at the module layer, then forwards the returned value back to the requester. The same mechanism is used for both single readback and repeated sampling.

As a concrete example, consider a request to read back the voltage of \texttt{MGTAVCC}. From Table~\ref{tab:kc705_lane_map}, \texttt{MGTAVCC} corresponds to lane 6, which maps to PMBus address 53 and \texttt{PAGE}=2. The resulting PMBus sequence is:

\begin{enumerate}
\item Write Byte: \\
\texttt{[Addr=53][PAGE (00h)][02h]}

\item Read Word: \\
\texttt{[Addr=53][READ\_VOUT (8Bh)]}
\end{enumerate}

Here, the PMBus address selects the regulator device and the PAGE command selects the target rail. The returned value is encoded in \texttt{LINEAR16} format and propagated back by the PowerManager as the measured rail voltage.

\subsubsection*{\textbf{How the UCD9248 interprets voltage setpoints}}

\begin{figure}[t]
\centering
\includegraphics[width=\linewidth]{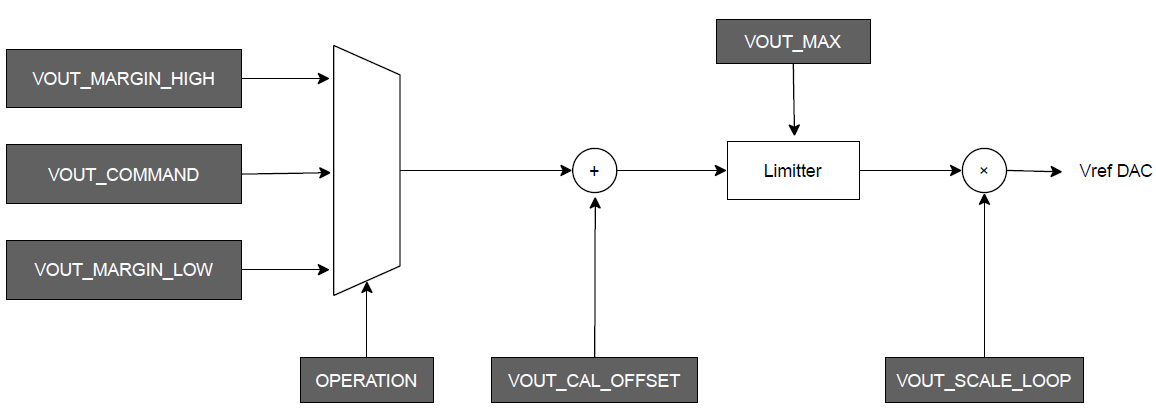}
\caption{UCD9248 voltage adjustment path.}
\label{fig:pmbus_vout_method}
\end{figure}

The UCD9248 does not apply \texttt{VOUT\_COMMAND} directly to the DAC. As shown in Figure~\ref{fig:pmbus_vout_method}, the programmed value is selected through the device control path, then combined with calibration offset, constrained by limits, and scaled before driving the DAC reference. This internal processing reinforces the point already made in Section~\ref{sec:architecture}: voltage adjustment must be treated as a regulator-level operation with finite response and settling time, not as an instantaneous rail change.

\subsection{Mapping Summary and Timing Discipline}
\label{sec:pmbus_summary_timing}

Table~\ref{tab:voltune_pmbus_map} summarizes how the main VolTune operations are implemented on KC705 using PMBus sequences and the transaction primitives introduced in Section~\ref{sec:pmbus_basics}. As discussed in Sections~\ref{sec:pmbus_rail_map} and \ref{sec:pmbus_sequences}, porting the same control approach to another PMBus-controlled multi-rail FPGA platform requires only the corresponding lane-to-(address,\texttt{PAGE}) mapping and the appropriate command sequence for each operation.

VolTune enforces serialized execution of PMBus transactions. A new PMBus request is not issued until the previous request completes at the module layer and returns completion status to the PowerManager. This avoids overlapping PMBus transactions and keeps regulator state unambiguous.

The end-to-end timing of a voltage operation is therefore determined by the transaction time on the PMBus wires together with regulator response and settling behavior. These timing components are quantified later in Section~\ref{sec:controller_eval}.

\begin{figure}[t]
\centering
\subfloat[1.0\,V $\rightarrow$ 0.5\,V transition (HW PMBus, 400\,kHz)]{\includegraphics[width=\linewidth]{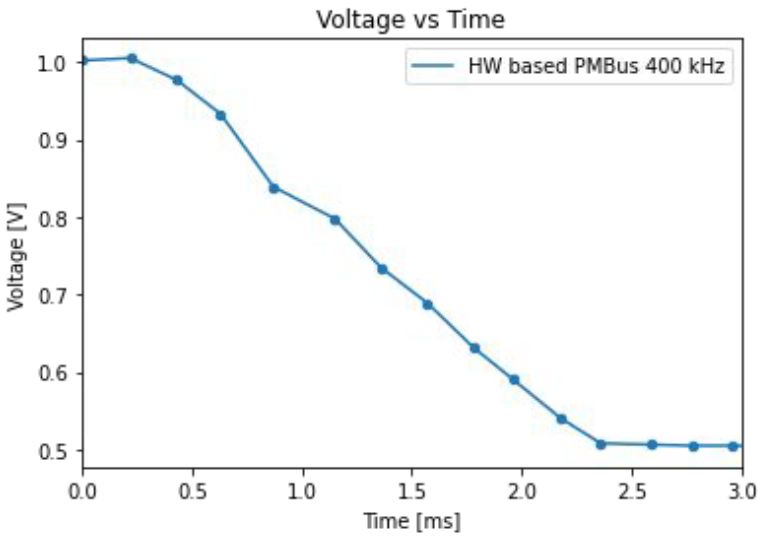}\label{fig:ctrl_measure_voltage_a}}\\
\vspace{0.6em}
\subfloat[Voltage transitions for multiple target voltages (HW PMBus, 400\,kHz)]{\includegraphics[width=\linewidth]{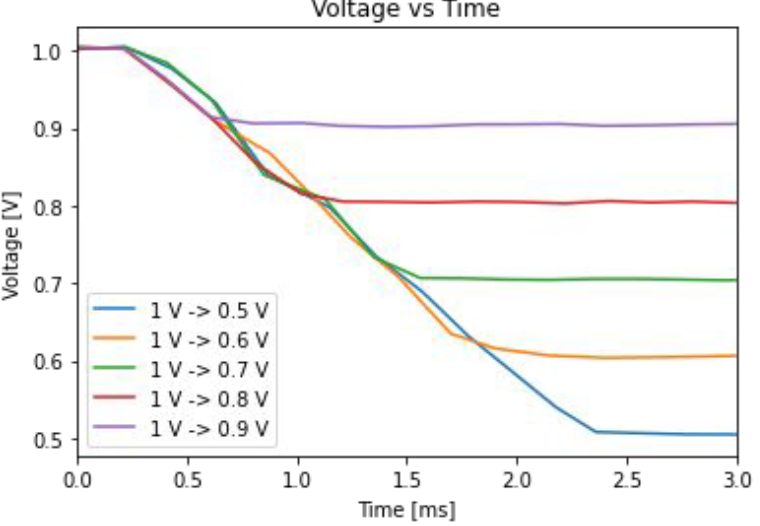}\label{fig:ctrl_measure_voltage_b}}
\caption{Measured voltage transition dynamics using the hardware PMBus control path at 400\,kHz.}
\label{fig:ctrl_measure_voltage}
\end{figure}
\begin{figure}[t]
\centering
\includegraphics[width=\linewidth]{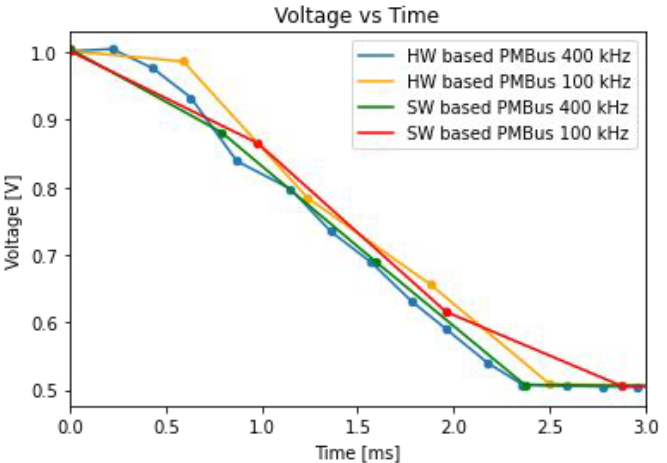}
\caption{Voltage transition comparison across control paths and PMBus clock rates.}
\label{fig:ctrl_voltage_compare}
\end{figure}
\begin{figure*}[!htbp]
\centering
\subfloat[Compute stable-voltage average $v_{\mathrm{avg}}$ from the last $N$ samples]{\includegraphics[width=0.48\linewidth]{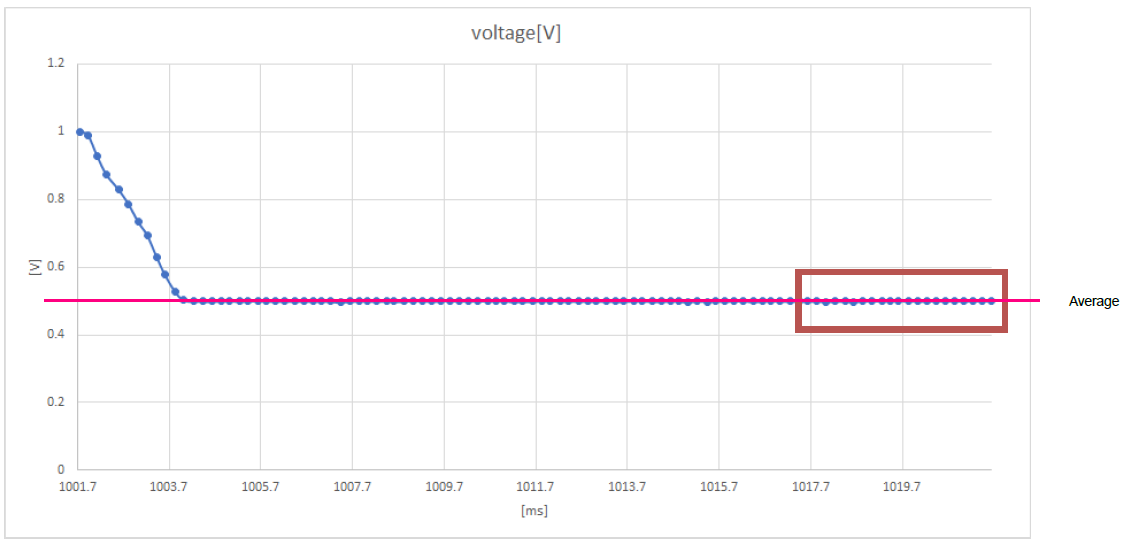}\label{fig:ctrl_settle_avg}}
\hfill
\subfloat[Define the stability band $v_{\mathrm{avg}}\pm x\%$]{\includegraphics[width=0.48\linewidth]{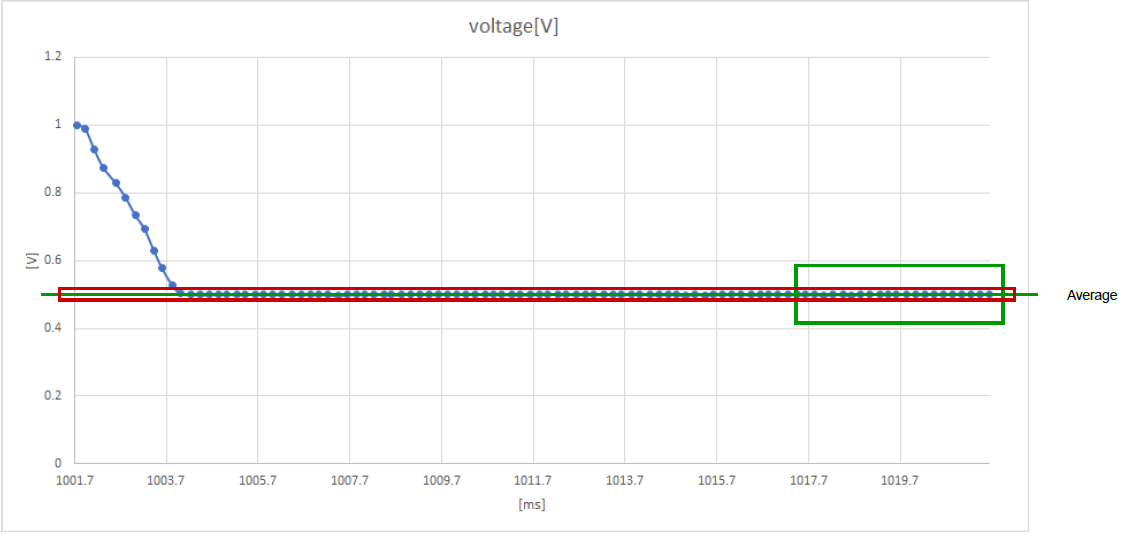}\label{fig:ctrl_settle_band}}\\
\vspace{0.6em}
\subfloat[Detect the first time $t_s$ with $N$ consecutive stable samples]{\includegraphics[width=0.48\linewidth]{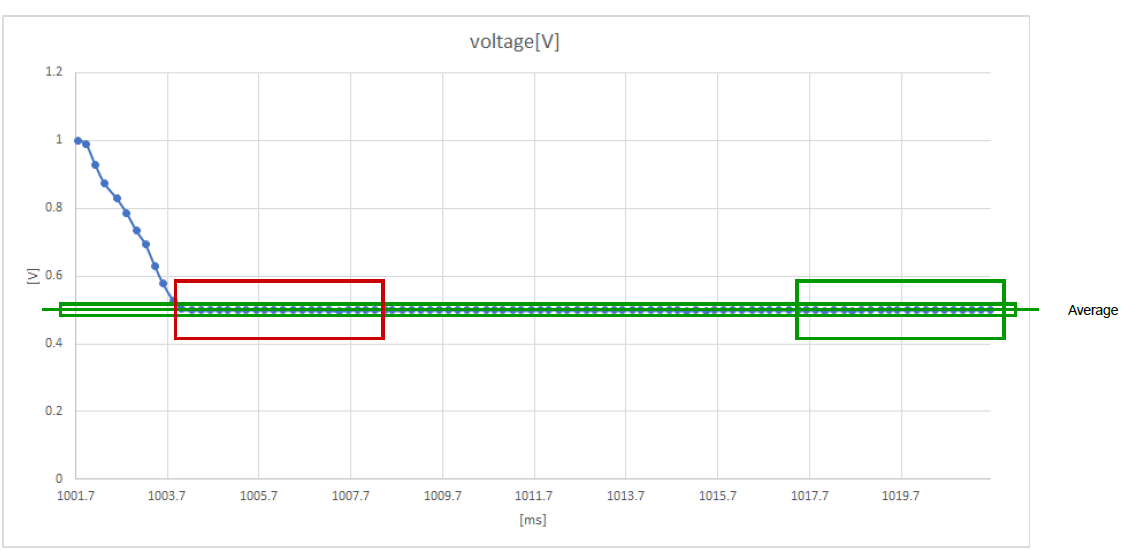}\label{fig:ctrl_settle_detect}}
\hfill
\subfloat[Settling time is the time from $t=0$ to $t_s$]{\includegraphics[width=0.48\linewidth]{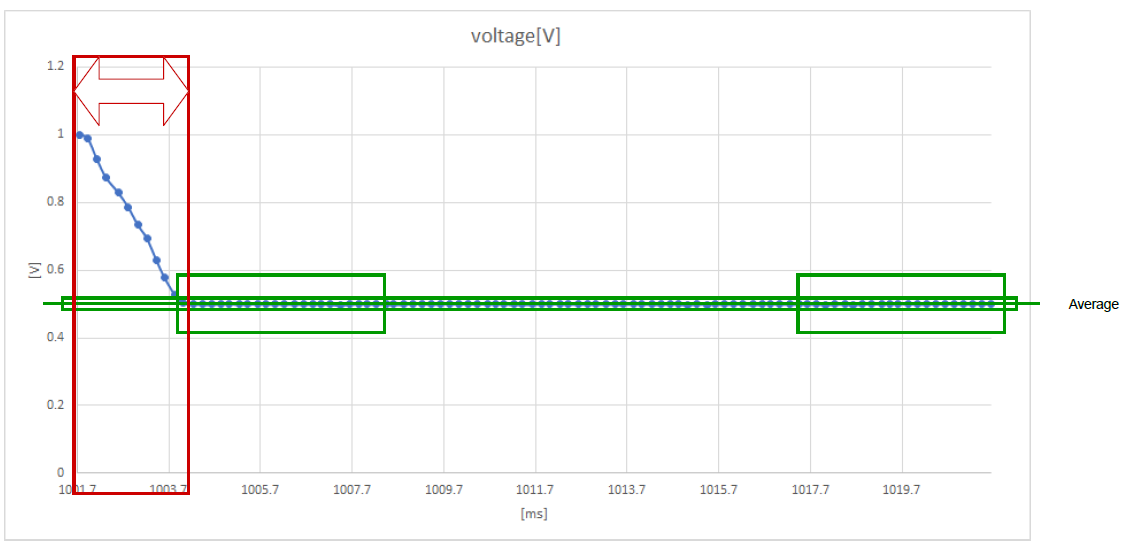}\label{fig:ctrl_settle_time}}
\caption{Settling-time detection method.}
\label{fig:ctrl_settling_method}
\end{figure*}
\section{Controller Characterization}
\label{sec:controller_eval}
In this section, we characterize the VolTune controller as a runtime voltage-control mechanism. We focus on voltage transition dynamics, control latency, settling behavior, and the impact of control-path choices and PMBus clocking. We also report controller overhead, including FPGA resource usage and static power, to quantify the cost of integrating runtime voltage control into the system. This controller-level evaluation is presented independently from the representative case study discussed later.

\subsection{Experimental Setup and Metrics}
\label{sec:ctrl_setup}

The controller characterization is performed on a KC705-based FPGA platform equipped with a TI UCD9248 digitally programmable power controller accessed through PMBus \cite{pmbus,ti_ucd9248}. We evaluate both VolTune control paths, the hardware control path implemented in FPGA logic and the software control path implemented on MicroBlaze, while keeping the same PMBus commands and rail-selection mechanism described in Section~\ref{sec:pmbus_impl}. As introduced in Section~\ref{sec:pmbus_iface}, the PMBus module is operated at two clock rates, 100\,kHz and 400\,kHz, which are evaluated here as implementation-level design choices affecting transaction timing. Unless stated otherwise, experiments target one supply rail at a time and apply controlled voltage transitions using VolTune opcodes translated into PMBus commands by the PowerManager.

Table~\ref{tab:ctrl_test_conditions} summarizes the settings used in this section. We sweep voltage transitions over multiple target voltages, evaluate both rising and falling transitions, and compare hardware versus software PMBus control paths under the two PMBus clock rates. This setup isolates controller behavior from application effects and enables direct comparison of transition dynamics and control overhead across configurations.

The characterization uses four metrics:
\begin{itemize}
    \item Voltage transition latency: the elapsed time from issuing a voltage-update request at the PowerManager interface to the point at which the measured rail voltage reaches and remains within a stable region around the target.
    \item Settling behavior: reported using voltage-versus-time traces and derived settling times under both rising and falling transitions.
    \item Control-path impact: evaluated by comparing hardware versus software execution and by varying the PMBus clock rate, which changes transaction overhead on SCL and SDA.
    \item Controller overhead: quantified using FPGA resource utilization and static power for both implementations.
\end{itemize}

\subsection{Voltage Transition Latency and Dynamics}
\label{sec:ctrl_latency}

Figure~\ref{fig:ctrl_measure_voltage} characterizes the voltage transition dynamics using the hardware PMBus control path at a 400\,kHz PMBus clock. We define voltage transition latency as the elapsed time from issuing a voltage-update request at the PowerManager interface to the point at which the measured rail voltage reaches the target and remains within a stable region around it. For the 1.0\,V $\rightarrow$ 0.5\,V transition in Figure~\ref{fig:ctrl_measure_voltage_a}, the measured end-to-end transition completes in 2.3\,ms.

Figure~\ref{fig:ctrl_measure_voltage_b} shows voltage trajectories for multiple target voltages. Under this hardware-PMBus 400\,kHz configuration, the results show a monotonic relationship between the voltage step size $\Delta V$ and the transition time. Larger steps, such as 1.0\,V $\rightarrow$ 0.5\,V, take longer than smaller steps, such as 1.0\,V $\rightarrow$ 0.9\,V. This behavior motivates the later comparison of rising and falling transitions across control-path implementations and PMBus clock rates.

\begin{table}[t]
\centering
\caption{Controller characterization test conditions.}
\label{tab:ctrl_test_conditions}
\begin{tabular}{l l}
\toprule
\textbf{Item} & \textbf{Values} \\
\midrule
PMBus control path  & Hardware-based PMBus\\
                    & Software-based PMBus \\
PMBus clock rate & 400 kHz, 100 kHz \\
Voltage decrease sweep & 1.0 V $\rightarrow$ 0.9, 0.8, 0.7, 0.6, 0.5 V \\
Voltage increase sweep & 0.5, 0.6, 0.7, 0.8, 0.9 V $\rightarrow$ 1.0 V \\
\bottomrule
\end{tabular}
\end{table}
\subsection{Impact of Control Path and PMBus Clock Rate}
\label{sec:ctrl_path_clock}

Figure~\ref{fig:ctrl_voltage_compare} compares voltage-transition measurements across four configurations: hardware-based PMBus at 400\,kHz and 100\,kHz, and software-based PMBus at 400\,kHz and 100\,kHz. The main effect of both the control path and the PMBus clock rate appears in the measurement interval during the transition, which is dominated by the time required to execute each PMBus transaction and collect a readback sample.

Table~\ref{tab:ctrl_sample_interval} summarizes the approximate interval reported for each configuration. The hardware-based control path achieves the finest measurement granularity, 0.2\,ms at 400\,kHz and 0.6\,ms at 100\,kHz. The software-based control path incurs substantially larger intervals, 0.8\,ms at 400\,kHz and 1.0\,ms at 100\,kHz, due to processor-side execution overhead in addition to the PMBus transaction time. These differences directly affect how quickly the controller can observe rail behavior and how densely it can sample the transient response, which later impacts settling-time detection and the repeatability of latency measurements.

\begin{table}[t]
\centering
\caption{Approximate measurement interval per configuration.}
\label{tab:ctrl_sample_interval}
\begin{tabular}{l c}
\toprule
\textbf{Configuration} & \textbf{Interval [ms]} \\
\midrule
HW-based PMBus, 400\,kHz & 0.2 \\
HW-based PMBus, 100\,kHz & 0.6 \\
SW-based PMBus, 400\,kHz & 0.8 \\
SW-based PMBus, 100\,kHz & 1.0 \\
\bottomrule
\end{tabular}
\end{table}
\subsection{Settling-Time Detection Methodology}
\label{sec:ctrl_settling_method}

To report controller latency consistently across configurations, we compute a settling time from the sampled voltage readback trace, following the procedure summarized in Figure~\ref{fig:ctrl_settling_method}. The method operates on a sequence of voltage samples $v[0],v[1],\dots,v[T]$ obtained during a transition. In Figure~\ref{fig:ctrl_settle_avg}, the stable voltage estimate is computed as the average of the last $N$ samples, $v_{\mathrm{avg}}=\frac{1}{N}\sum_{k=T-N+1}^{T} v[k]$. Figure~\ref{fig:ctrl_settle_band} then defines a stability band around this value as $v_{\mathrm{avg}}\pm x\%$, and samples within this band are classified as stable. The next step, illustrated in Figure~\ref{fig:ctrl_settle_detect}, is to locate the first time index $t_s$ such that $N$ consecutive samples beginning at $t_s$ are stable. Finally, Figure~\ref{fig:ctrl_settle_time} defines the settling time as the elapsed time from the first sample ($t=0$) to $t_s$. This criterion makes the definition robust to transient overshoot and measurement noise, while remaining simple to reproduce across different PMBus clock rates and control-path implementations.

\subsection{Validation Against External Measurement}
\label{sec:ctrl_validation}
\begin{figure*}[!htbp]
\centering
\subfloat[PMBus-based voltage readback]{\includegraphics[width=0.48\linewidth]{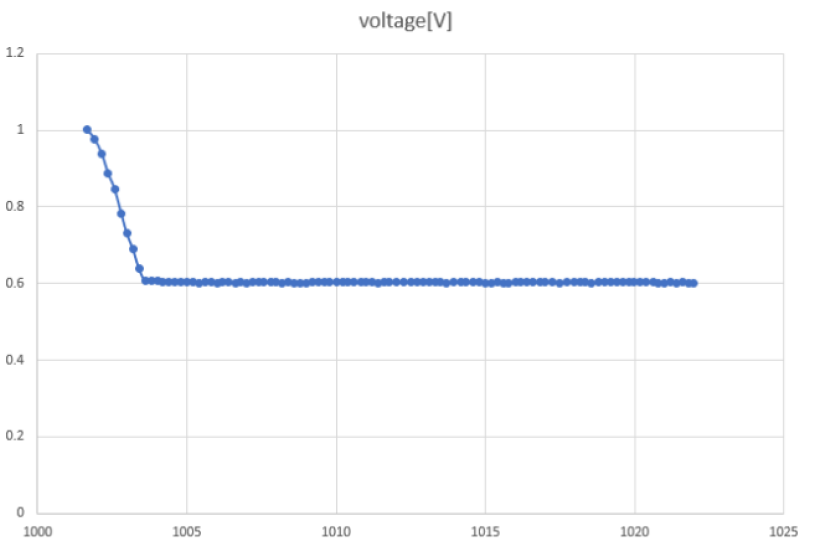}\label{fig:ctrl_validation_pmbus}}
\hfill
\subfloat[Oscilloscope waveform]{\includegraphics[width=0.48\linewidth]{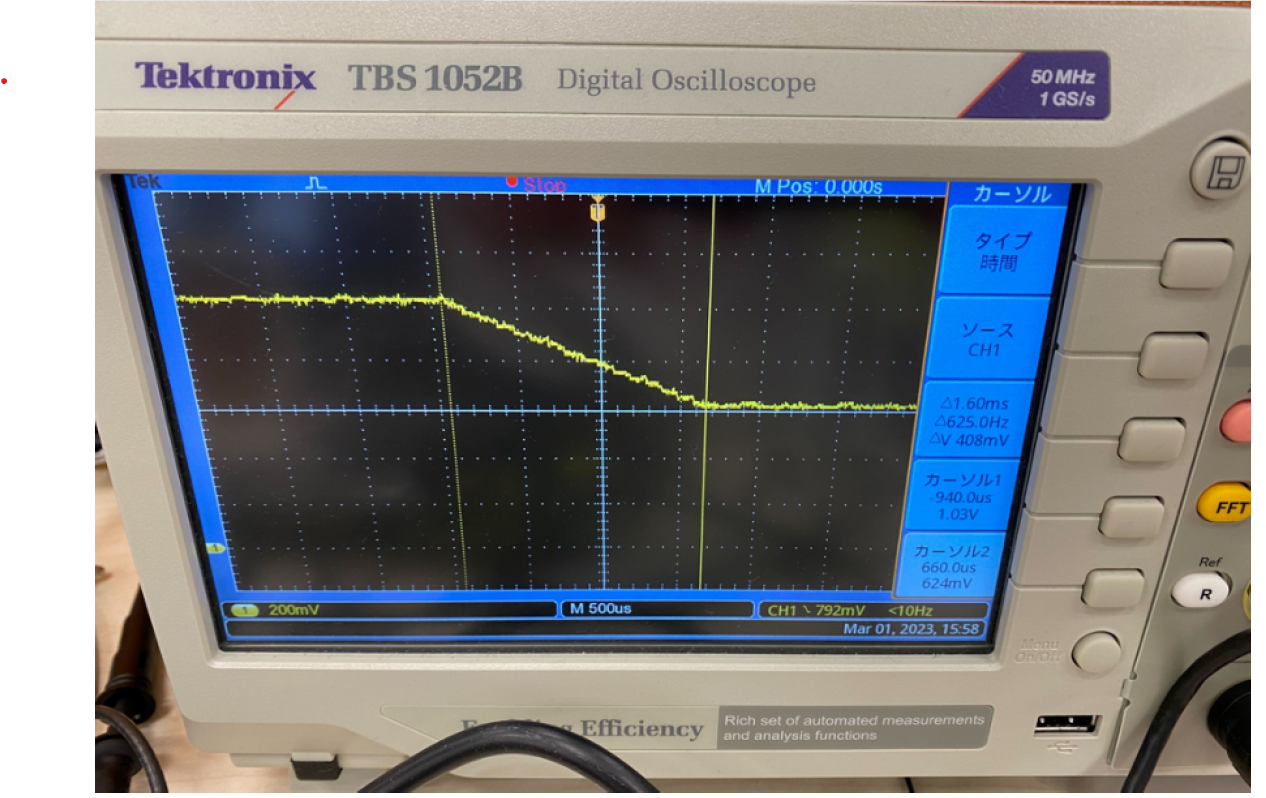}\label{fig:ctrl_validation_scope}}
\caption{PMBus readback and oscilloscope validation.}
\label{fig:ctrl_validation}
\end{figure*}

To validate the internal PMBus-based voltage readback measurements used throughout this section, we compare them against an external oscilloscope measurement of the same voltage transition. Figure~\ref{fig:ctrl_validation_pmbus} shows the voltage trace obtained from periodic PMBus readback using \texttt{READ\_VOUT}, while Figure~\ref{fig:ctrl_validation_scope} shows the oscilloscope waveform captured at the same rail.

The two measurements reach the same target voltage and show the same overall transition trend. A visible timing offset can occur because the PMBus-based trace is derived from discrete samples taken at the measurement interval determined by the selected control path and PMBus clock rate, whereas the oscilloscope provides higher-time-resolution analog sampling. In the remainder of this section, we use the PMBus-based readback for consistency across configurations and apply the settling-time detection methodology described in Section~\ref{sec:ctrl_settling_method} to infer transition latency from sampled observations.
\begin{table*}[t]
\centering
\caption{Hardware-based implementation resource utilization.}
\label{tab:ctrl_resource_breakdown_hw}
\begin{tabular}{l c c c c c c c c}
\toprule
\textbf{Component/Resource} &
\textbf{Slice LUTs} &
\textbf{Slice Reg.} &
\textbf{Slices} &
\textbf{LUT as Logic} &
\textbf{LUT as Mem.} &
\textbf{BRAM Tile} &
\textbf{DSPs} \\
&
\textbf{(203800)} &
\textbf{(407600)} &
\textbf{(50950)} &
\textbf{(203800)} &
\textbf{(64000)} &
\textbf{(445)} &
\textbf{(840)} \\
\midrule
\texttt{Counter module}& $<$0.01\% & 0.02\% & 0.03\% & $<$0.01\% & 0.00\% & 0.00\% & 0.00\% \\
\texttt{PowerManager module}& 0.31\% & 0.46\% & 1.19\% & 0.31\% & 0.02\% & 0.00\% & 0.24\% \\
\texttt{PMBus module} & 0.12\% & 0.03\% & 0.15\% & 0.12\% & 0.00\% & 0.00\% & 0.00\% \\
\texttt{Total} & 1.45\% & 1.30\% & 3.48\% & 1.22\% & 0.72\% & 1.80\% & 0.24\% \\
\bottomrule
\end{tabular}
\end{table*}

\begin{table*}[t]
\centering
\caption{Software-based implementation resource utilization.}
\label{tab:ctrl_resource_breakdown_sw}
\begin{tabular}{l c c c c c c c c}
\toprule
\textbf{Component/Resource} &
\textbf{Slice LUTs} &
\textbf{Slice Reg.} &
\textbf{Slices} &
\textbf{LUT as Logic} &
\textbf{LUT as Mem.} &
\textbf{BRAM Tile} &
\textbf{DSPs} \\
&
\textbf{(203800)} &
\textbf{(407600)} &
\textbf{(50950)} &
\textbf{(203800)} &
\textbf{(64000)} &
\textbf{(445)} &
\textbf{(840)} \\
\midrule
\texttt{axi\_gpio}& 0.03\% & 0.02\% & 0.05\% & 0.03\% & 0.00\% & 0.00\% & 0.00\% \\
\texttt{axi\_timer}& 0.10\% & 0.04\% & 0.16\% & 0.10\% & 0.00\% & 0.00\% & 0.00\% \\
\texttt{axi\_uartlite}& 0.05\% & 0.03\% & 0.09\% & 0.04\% & 0.02\% & 0.00\% & 0.00\% \\
\texttt{axis\_dwidth\_converter}& $<$0.01\% & 0.06\% & 0.11\% & $<$0.01\% & 0.00\% & 0.00\% & 0.00\% \\
\texttt{mdm\_1}& 0.05\% & 0.03\% & 0.08\% & 0.04\% & 0.01\% & 0.00\% & 0.00\% \\
\texttt{microblaze} & 0.76\% & 0.31\% & 1.12\% & 0.70\% & 0.19\% & 0.00\% & 0.36\% \\
\texttt{microblaze\_local\_memory}& 0.36\% & 0.32\% & 0.98\% & 0.24\% & 0.40\% & 57.53\% & 0.00\% \\
\texttt{pmbus\_io}& 0.00\% & 0.00\% & 0.00\% & 0.00\% & 0.00\% & 0.00\% & 0.00\% \\
\texttt{smartconnect}& 0.19\% & 0.09\% & 0.36\% & 0.19\% & $<$0.01\% & 0.00\% & 0.00\% \\
\texttt{util\_vector\_logic}& $<$0.01\% & 0.00\% & $<$0.01\% & $<$0.01\% & 0.00\% & 0.00\% & 0.00\% \\
\texttt{Total}& 1.53\% & 0.90\% & 2.81\% & 1.34\% & 0.62\% & 57.52\% & 0.36\% \\
\bottomrule
\end{tabular}
\end{table*}
\subsection{Controller Overhead}
\label{sec:ctrl_overhead}
We quantify the overhead of integrating VolTune by reporting FPGA resource utilization and estimated static power for the hardware-based and software-based implementations. Resource utilization is extracted from Vivado implementation reports, and static power is extracted from Vivado power reports. The hardware-based implementation realizes the PowerManager subsystem and PMBus transaction engine as dedicated FPGA logic, as shown in Figure~\ref{fig:powermanager_arch}, while the software-based implementation additionally instantiates a MicroBlaze processor subsystem together with its associated memory and interconnect, as shown in Figure~\ref{fig:sw_powermanager_bd}.

\subsubsection*{\textbf{FPGA resource utilization}}

Table~\ref{tab:ctrl_resource_breakdown_hw} and Table~\ref{tab:ctrl_resource_breakdown_sw} report the Vivado hierarchical utilization breakdown for the hardware-based and software-based VolTune implementations, respectively. Percentages are shown relative to the KC705 device totals listed in the column headers.

After excluding debug logic, the main overhead difference between the two implementations is not uniform across all resource classes. The software-based implementation remains close to the hardware-based implementation in Slice LUTs, 1.53\% versus 1.45\%, and increases DSP usage from 0.24\% to 0.36\% (1.50$\times$). At the same time, its Slice Register and Slice utilization percentages are lower than those of the hardware-based implementation, 0.90\% versus 1.30\% for Slice Registers and 2.81\% versus 3.48\% for Slices. The dominant difference is Block RAM utilization, which increases from 1.80\% to 57.52\% (31.96$\times$), indicating that the primary cost of the software control path is the processor memory subsystem rather than additional PMBus-specific logic.

The breakdown also shows where the overhead originates. In the software-based design, \texttt{microblaze\_local\_memory} alone consumes 57.53\% of BRAM tiles, accounting for essentially all of the total BRAM usage. In contrast, the remaining infrastructure, including \texttt{smartconnect} (0.19\% LUTs, 0.09\% registers, 0.36\% slices) and AXI peripherals such as \texttt{axi\_timer} and \texttt{axi\_uartlite}, contributes relatively small fractions. This indicates that the software implementation overhead is concentrated in instantiating a general-purpose execution substrate rather than in scaling the PMBus transaction engine itself.

For the hardware-based implementation, the PMBus-specific modules remain lightweight. The \texttt{PMBus module} consumes 0.12\% Slice LUTs and 0.15\% slices with no BRAM usage, while the \texttt{PowerManager module} consumes 0.31\% Slice LUTs and 1.19\% slices with no BRAM usage and 0.24\% DSP usage. The \texttt{Counter module} is negligible across all resources. As a result, the hardware control path provides a substantially lower-memory integration point, and its overhead is dominated by standard FPGA logic resources rather than BRAM, which is often the limiting resource for accelerator designs.

Overall, these results support the design choice of providing both control paths. The hardware-based control path offers the lower-memory and lower-static-power implementation, while the software-based control path preserves comparable logic utilization but trades a 31.96$\times$ increase in BRAM utilization for programmability and ease of extension.

\subsubsection*{\textbf{Static power estimation}}
Table~\ref{tab:ctrl_static_power_breakdown} reports the Vivado static power breakdown for the hardware-based and software-based VolTune implementations. A direct comparison shows that the software-based control path increases static power from 0.015\,W to 0.084\,W (5.60$\times$), reflecting the always-on processor and memory infrastructure required by the MicroBlaze-based execution substrate.

\begin{table}[t]
\centering
\caption{VolTune static power breakdown.}
\label{tab:ctrl_static_power_breakdown}
\setlength{\tabcolsep}{4pt}
\renewcommand{\arraystretch}{1.05}
\begin{tabular}{l l r l}
\toprule
\textbf{Implementation} & \textbf{Module} & \textbf{W} & \textbf{Share} \\
\midrule
\multirow{4}{*}{Hardware}
  & \texttt{PowerManager module}  & 0.011 & 1\% \\
  & \texttt{PMBus module} & 0.003 & 1\% \\
  & \texttt{Counter module}       & 0.001 & $<$1\% \\
  \cmidrule(lr){2-4}
  & \textbf{Total}                & \textbf{0.015} & \textbf{2\%} \\
\midrule
\multirow{9}{*}{Software}
  & \texttt{microblaze}                & 0.052 & 6\% \\
  & \texttt{microblaze\_local\_memory} & 0.023 & 3\% \\
  & \texttt{smartconnect}              & 0.003 & 1\% \\
  & \texttt{axi\_timer}                & 0.002 & $<$1\% \\
  & \texttt{axis\_dwidth\_converter}   & 0.001 & $<$1\% \\
  & \texttt{axi\_uartlite}             & 0.001 & $<$1\% \\
  & \texttt{mdm\_1}                    & 0.001 & $<$1\% \\
  & \texttt{axi\_gpio}                 & 0.001 & $<$1\% \\
  \cmidrule(lr){2-4}
  & \textbf{Total}                     & \textbf{0.084} & \textbf{9\%} \\
\bottomrule
\end{tabular}
\end{table}

The breakdown indicates that the software-based overhead is dominated by the processor subsystem. The \texttt{microblaze} core (0.052\,W) and \texttt{microblaze\_local\_memory} (0.023\,W) contribute 0.075\,W combined, accounting for 89.3\% of the 0.084\,W total. In contrast, the remaining AXI peripherals and interconnect blocks each contribute 0.003\,W or less, indicating that the static-power overhead is concentrated in the general-purpose execution substrate rather than in peripheral glue logic.

For the hardware-based implementation, the dominant contributor is the \texttt{PowerManager module} at 0.011\,W, while the remaining support modules each contribute 0.003\,W or less. This confirms that the hardware control path has a very small always-on static-power footprint compared with the software-based alternative.

\subsection{Summary}
\label{sec:ctrl_summary}

Overall, the controller characterization shows that the hardware-based VolTune control path has a low integration cost in both power and FPGA resources. Its always-on static-power overhead remains around only a few percent of the subsystem budget, and its resource usage occupies only a small fraction of the KC705 device. These results indicate that VolTune can operate as a practical runtime control utility block without materially reducing the power or resource budget available to the main application logic.


\section{Representative Case Study}
\label{sec:case_study}

\subsection{Case-Study Target and Rationale}
\label{sec:cs_target}

We evaluate VolTune on the KC705 high-speed serial transceivers as a representative, voltage-sensitive subsystem. The case study targets the transceiver supply rails generated by the Texas Instruments UCD9248 power controller and therefore directly programmable through the PMBus-controlled regulator \cite{ug810_kc705}.

Serial transceivers are a practical target because voltage tuning can be evaluated using application-visible metrics under multi-gigabit operation, primarily bit error rate (BER) and also link latency. In addition, transceiver rails are largely decoupled from core logic supplies, which enables targeted rail-level experiments without modifying the FPGA bitstream or the timing closure of unrelated logic.

The results in this section are intended as a representative demonstration. The objective is to show that VolTune can be used to sweep rail voltages at runtime, quantify energy and reliability behavior, and identify a bounded operating region on a commercial FPGA platform.

\subsection{Experimental Setup and Methodology}
\label{sec:cs_setup}

This case study evaluates runtime voltage tuning on the KC705 transceiver subsystem through repeatable link tests. The evaluation loop programs the target rail voltage through VolTune and measures link correctness and performance at each operating point. Voltage updates and readback reuse the same VolTune opcode and PMBus control mechanisms described in Section~\ref{sec:pmbus_impl}.

\subsubsection*{\textbf{Board setup}}
\begin{figure}[t]
\centering
\includegraphics[width=\linewidth]{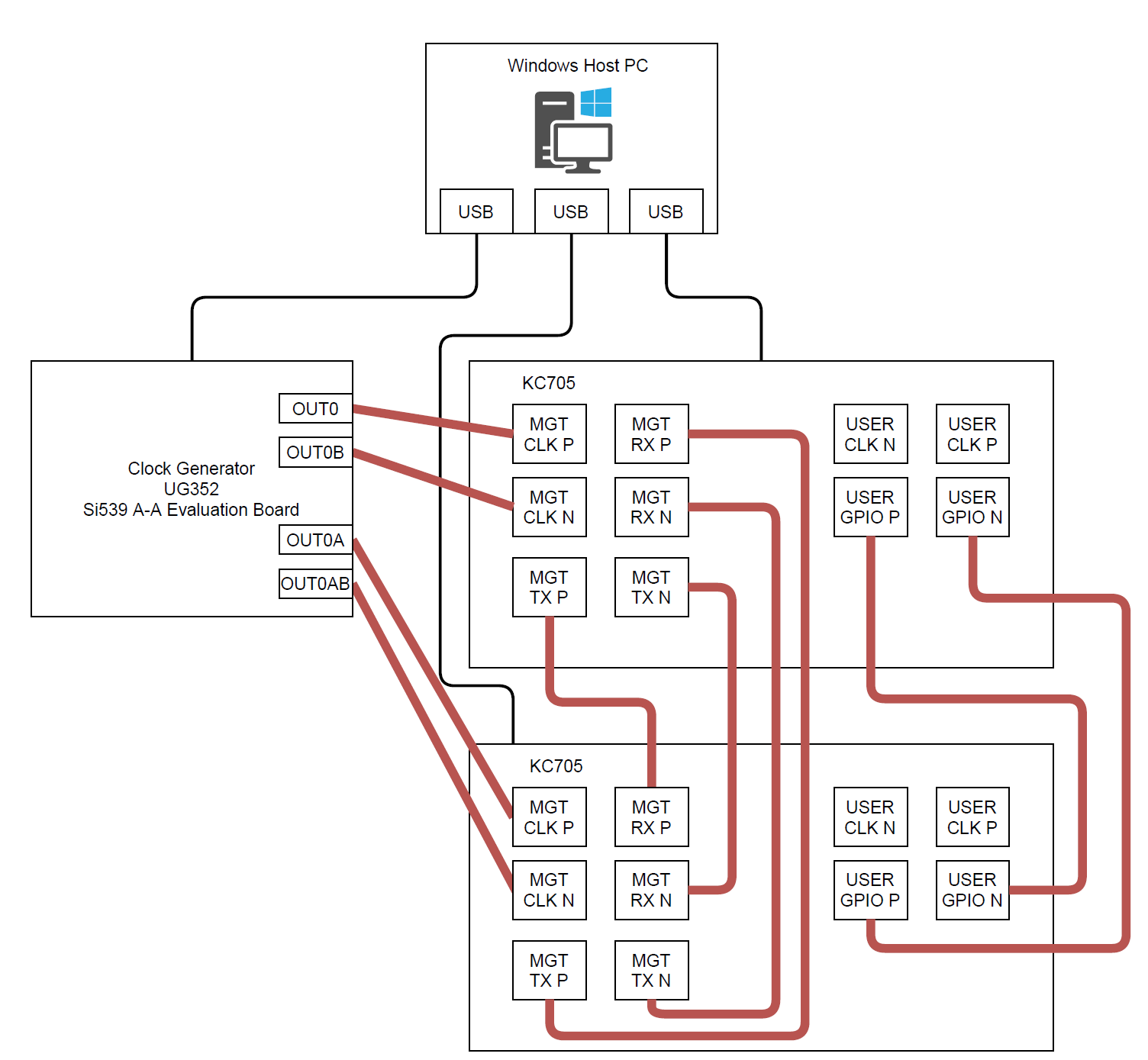}
\caption{Board setup for the transceiver case study.}
\label{fig:cs_board_setup}
\end{figure}

Figure~\ref{fig:cs_board_setup} shows the board-level setup used for the transceiver case study. The evaluation uses two KC705 boards connected back-to-back through the on-board high-speed transceiver links. A Windows Host PC runs the test program and controls both boards through JTAG using \texttt{hw\_server} and XSDB. A dedicated external clock generator, the Skyworks Si5391A-A evaluation board \cite{ug352_si5391a_evb}, provides the reference clocks used by the transceivers on both KC705 boards. In our setup, the clock generator outputs 125.000\,MHz for 2.5, 5.0, and 10.0\,Gbps tests, and 117.188\,MHz for 7.5\,Gbps tests, as summarized in Table~\ref{tab:cs_test_conditions}. The selected reference clock is routed to the corresponding transceiver reference clock inputs of both boards, ensuring that both ends of the link operate from a consistent and repeatable reference clock during each voltage sweep.

\subsubsection*{\textbf{Target voltage rail}}
We evaluate VolTune in a representative setting based on the GTX transceivers on KC705 \cite{amd_ug476_gtx}. In this case study, voltage tuning is applied to the GTX transceiver power domain by sweeping the analog supply rail \texttt{MGTAVCC}, previously listed in Table~\ref{tab:kc705_lane_map}. For Kintex-7 devices, \texttt{MGTAVCC} supplies the GTX transmitter and receiver analog circuits, while \texttt{MGTAVTT} supplies the termination circuitry \cite{amd_ds182_kintex7,amd_ug476_gtx}. In our setup, \texttt{MGTAVTT} is held at its nominal configuration while \texttt{MGTAVCC} is swept, to isolate the effect of analog supply scaling.

\subsubsection*{\textbf{Workload and sweep procedure}}
For each voltage point, the transmitter sends a 10\,GByte count-up data stream and the receiver checks correctness to measure received data size and bit error rate (BER). Tests are repeated across link speeds of 2.5, 5.0, 7.5, and 10.0\,Gbps. To separate transmitter-side and receiver-side sensitivity, we sweep the transceiver supply rail for the TX side and the RX side independently using three modes: (i) TX and RX swept together, (ii) TX fixed at nominal 1.0\,V with RX swept, and (iii) RX fixed at nominal 1.0\,V with TX swept. Table~\ref{tab:cs_test_conditions} summarizes the full set of test conditions used in this case study.
\begin{table}[t]
\centering
\caption{Case-study test conditions.}
\label{tab:cs_test_conditions}
\setlength{\tabcolsep}{6pt}
\renewcommand{\arraystretch}{1.05}
\begin{tabular}{l l}
\toprule
\textbf{Item} & \textbf{Value} \\
\midrule
Target rails &
\begin{tabular}[t]{@{}l@{}}
\texttt{MGTAVCC} swept \\
\texttt{MGTAVTT} fixed at nominal
\end{tabular} \\
\midrule
Voltage sweep &
\begin{tabular}[t]{@{}l@{}}
Range: 1.0\,V $\rightarrow$ 0.7\,V \\
Step: 0.001\,V
\end{tabular} \\
\midrule
Sweep modes &
\begin{tabular}[t]{@{}l@{}}
TX and RX swept together \\
TX fixed at 1.0\,V, RX swept \\
RX fixed at 1.0\,V, TX swept
\end{tabular} \\
\midrule
Link speed (reference clock) &
\begin{tabular}[t]{@{}l@{}}
2.5\,Gbps (125.000\,MHz) \\
5.0\,Gbps (125.000\,MHz) \\
7.5\,Gbps (117.188\,MHz) \\
10.0\,Gbps (125.000\,MHz)
\end{tabular} \\
\midrule
Test payload &
\begin{tabular}[t]{@{}l@{}}
10\,GByte count-up data stream
\end{tabular} \\
\midrule
PMBus control path &
\begin{tabular}[t]{@{}l@{}}
Hardware-based PMBus
\end{tabular} \\
\midrule
PMBus clock rate &
\begin{tabular}[t]{@{}l@{}}
400\,kHz
\end{tabular} \\
\midrule
Reported metrics &
\begin{tabular}[t]{@{}l@{}}
Received data size \\
Bit error rate (BER) \\
Latency \\
Rail power
\end{tabular} \\
\bottomrule
\end{tabular}
\end{table}

\subsection{Bit Error Rate and Throughput Under Voltage Tuning}
\label{sec:cs_ber}
\begin{figure*}[t]
\centering
\subfloat[Received data size]{\includegraphics[width=0.32\linewidth]{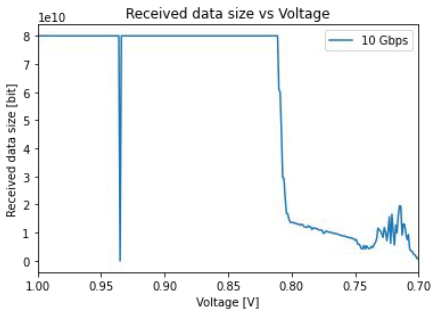}\label{fig:cs_recvsize_10g}}
\hfill
\subfloat[BER over full sweep]{\includegraphics[width=0.32\linewidth]{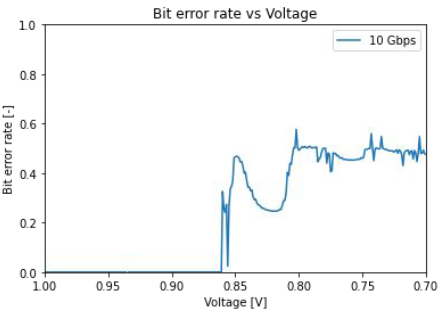}\label{fig:cs_ber_10g}}
\hfill
\subfloat[BER near the onset region (log scale)]{\includegraphics[width=0.32\linewidth]{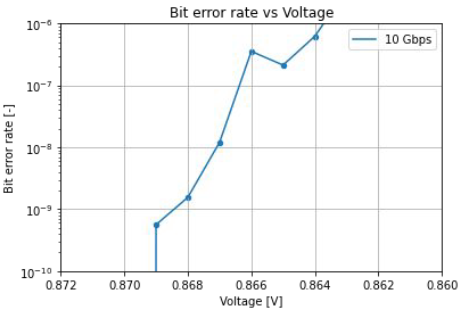}\label{fig:cs_ber_10g_zoom}}
\caption{10.0\,Gbps reliability under voltage tuning.}
\label{fig:cs_ber_recvsize_10g}
\end{figure*}

We first evaluate how link reliability changes when the transceiver analog supply rail is reduced at runtime. We use the 10.0\,Gbps link setting and sweep the target rail from 1.0\,V down to 0.7\,V using the procedure described in Section~\ref{sec:cs_setup}. For each voltage point, the transmitter sends a 10\,GByte count-up stream and the receiver reports both the received data size and the bit error rate (BER). Received data size is included to distinguish gradual BER degradation from hard link failure or loss of synchronization.

Figure~\ref{fig:cs_ber_recvsize_10g} reveals three regimes that are relevant to runtime voltage selection. First, the link delivers the full payload and BER remains effectively zero down to approximately 0.869\,V, as shown by the flat region in Figure~\ref{fig:cs_ber_10g}. Second, errors begin to appear at approximately 0.869\,V, and BER increases from the $10^{-10}$--$10^{-9}$ range near 0.869--0.868\,V to approximately $10^{-7}$ near 0.866\,V, and to approximately $10^{-6}$ near 0.864\,V in Figure~\ref{fig:cs_ber_10g_zoom}. Third, at lower voltages the link becomes unstable and the received data size drops sharply, with the first major throughput collapse occurring near 0.80\,V in Figure~\ref{fig:cs_recvsize_10g}.

These observations indicate that runtime tuning can operate within a bounded usable region above the collapse point, and that the voltage range near 0.869--0.864\,V forms a transition region where BER increases by several orders of magnitude without an immediate loss of link activity. In the following subsections, we use these regimes as a reference point to separate TX-side and RX-side sensitivity and to relate error onset to latency and rail power.

\subsection{Sensitivity Study, TX-Only vs RX-Only Voltage Scaling}
\label{sec:cs_tx_rx}
\begin{figure*}[t]
\centering
\subfloat[Received data size]{\includegraphics[width=0.44\linewidth]{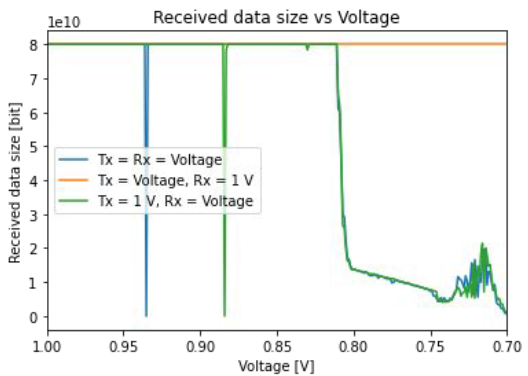}\label{fig:cs_txrx_recvsize_cmp}}
\hfill
\subfloat[BER (full sweep)]{\includegraphics[width=0.44\linewidth]{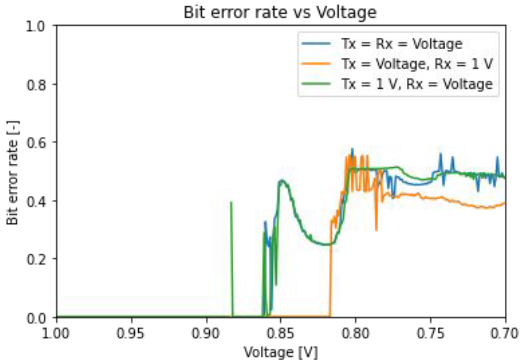}\label{fig:cs_txrx_ber_cmp}}
\caption{TX-only vs RX-only voltage scaling at 10.0\,Gbps.}
\label{fig:cs_txrx_recv_ber}
\end{figure*}

We next separate TX-side and RX-side voltage sensitivity by comparing three sweep modes: (i) TX and RX swept together, (ii) TX swept while RX is fixed at 1.0\,V, and (iii) RX swept while TX is fixed at 1.0\,V. Figure~\ref{fig:cs_txrx_recv_ber} summarizes the received data size and BER trends for the three modes at 10.0\,Gbps.

Two quantitative differences stand out. First, when RX is fixed at 1.0\,V and only TX voltage is swept, the received data size remains at the full payload down to 0.7\,V. In contrast, when RX voltage is swept (either TX=RX swept together or TX fixed at 1.0\,V), the received data size begins to decrease at approximately 0.81\,V (Figure~\ref{fig:cs_txrx_recvsize_cmp}). Second, the BER onset voltage shifts substantially depending on whether RX is swept. When RX is fixed, BER does not increase until approximately 0.82\,V, while in the RX-swept cases BER begins to increase at approximately 0.87\,V. These onset voltages are derived from the corresponding zoomed BER analysis for the same measurements, which is omitted here for brevity.

Overall, these results indicate that RX-side voltage reduction has a larger impact on both throughput stability and BER than TX-side voltage reduction in this setup. In the following subsections, we use this asymmetry to interpret how speed and power trends evolve under voltage tuning.

\subsection{Impact of Link Speed}
\label{sec:cs_speed}
\begin{figure*}[t]
\centering
\subfloat[Received data size]{\includegraphics[width=0.48\linewidth]{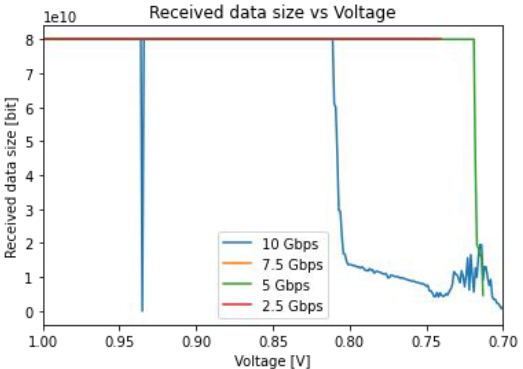}\label{fig:cs_speed_recvsize}}
\hfill
\subfloat[BER (full sweep)]{\includegraphics[width=0.48\linewidth]{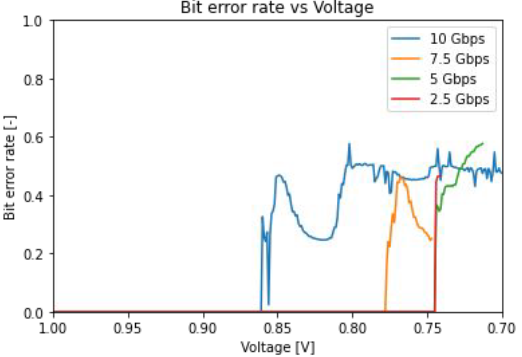}\label{fig:cs_speed_ber}}
\caption{Link-speed impact under voltage tuning.}
\label{fig:cs_speed_full}
\end{figure*}

We next evaluate how the voltage sensitivity shifts with transceiver link speed. Figure~\ref{fig:cs_speed_full} compares received data size and BER over the full voltage sweep for 2.5, 5.0, 7.5, and 10.0\,Gbps under the same test methodology in Section~\ref{sec:cs_setup}.

Figure~\ref{fig:cs_speed_full} shows that higher link speed requires higher voltage to maintain low-error operation. For 10.0\,Gbps, BER begins to rise at approximately 0.869\,V. For 7.5\,Gbps, the BER onset shifts down to approximately 0.787\,V. For 5.0\,Gbps and 2.5\,Gbps, the BER onset occurs at approximately 0.745\,V and 0.744\,V, respectively. These onset voltages indicate a widening voltage headroom as link speed decreases, which increases the available tuning range for reducing rail power.

Received data size provides a complementary view of stability. In Figure~\ref{fig:cs_speed_recvsize}, the 10.0\,Gbps case exhibits the earliest throughput collapse, with a major drop in received data size near 0.80\,V. In contrast, the 5.0\,Gbps case maintains full received data size to a lower voltage and then drops near 0.72\,V. The 7.5\,Gbps and 2.5\,Gbps tests terminate before a clear received-data-size collapse is observed in the plotted sweep, so BER onset provides the primary stability indicator for these two speeds in this dataset.

Overall, these results show that the usable voltage range depends strongly on link speed. The transition region shifts downward as speed decreases, enabling more aggressive voltage reduction at lower rates while preserving bounded BER and stable throughput.

\subsection{Latency Impact}
\label{sec:cs_latency}
\begin{figure*}[t]
\centering
\subfloat[TX-only vs RX-only scaling modes]{\includegraphics[width=0.48\linewidth]{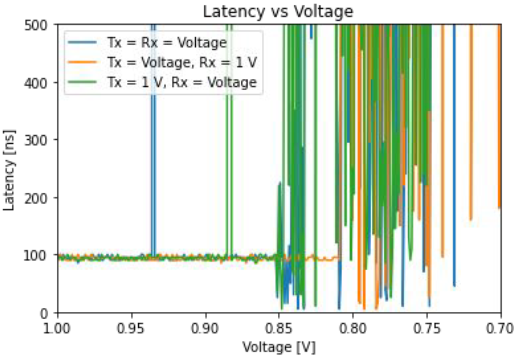}\label{fig:cs_latency_modes}}
\hfill
\subfloat[Link-speed comparison]{\includegraphics[width=0.48\linewidth]{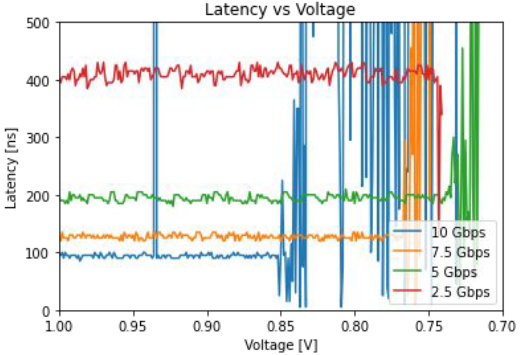}\label{fig:cs_latency_speeds}}
\caption{Latency under voltage tuning.}
\label{fig:cs_latency}
\end{figure*}

We next evaluate how voltage tuning impacts end-to-end link latency. Figure~\ref{fig:cs_latency} reports measured latency as a function of swept voltage and link speed. We present two complementary views: a comparison across TX-only versus RX-only scaling modes (Figure~\ref{fig:cs_latency_modes}) and a comparison across link speeds (Figure~\ref{fig:cs_latency_speeds}).
\begin{table*}[t]
\centering
\caption{TX-side and RX-side power trends under different sweep modes.}
\label{tab:cs_power_mode_summary}
\setlength{\tabcolsep}{5pt}
\renewcommand{\arraystretch}{1.05}
\begin{tabular}{l c c}
\toprule
\textbf{Sweep mode} & \textbf{TX-side power trend} & \textbf{RX-side power trend} \\
\midrule
TX = RX swept &
Decreases from $\sim$0.20\,W at 1.0\,V to $\sim$0.08\,W at 0.7\,V &
Decreases from $\sim$0.17\,W at 1.0\,V to $\sim$0.07--0.08\,W at 0.7\,V \\
\midrule
TX swept, RX fixed at 1.0\,V &
Decreases from $\sim$0.20\,W at 1.0\,V to $\sim$0.08\,W at 0.7\,V &
Approximately constant near $\sim$0.16--0.17\,W \\
\midrule
RX swept, TX fixed at 1.0\,V &
Approximately constant near $\sim$0.20\,W &
Decreases from $\sim$0.17\,W at 1.0\,V to $\sim$0.07--0.08\,W at 0.7\,V \\
\bottomrule
\end{tabular}
\end{table*}

\begin{table*}[t]
\centering
\caption{Representative rail power versus link speed.}
\label{tab:cs_power_speed_summary}
\setlength{\tabcolsep}{5pt}
\renewcommand{\arraystretch}{1.05}
\begin{tabular}{l c c c c}
\toprule
\textbf{Speed} &
\textbf{TX power @ 1.0\,V} &
\textbf{RX power @ 1.0\,V} &
\textbf{TX power @ 0.8\,V} &
\textbf{RX power @ 0.8\,V} \\
\midrule
10.0\,Gbps & $\sim$0.20\,W & $\sim$0.17\,W & $\sim$0.13\,W & $\sim$0.11\,W \\
7.5\,Gbps  & $\sim$0.18\,W & $\sim$0.15--0.16\,W & $\sim$0.12\,W & $\sim$0.10\,W \\
5.0\,Gbps  & $\sim$0.14\,W & $\sim$0.12\,W & $\sim$0.09\,W & $\sim$0.08\,W \\
2.5\,Gbps  & $\sim$0.12\,W & $\sim$0.09--0.10\,W & $\sim$0.08\,W & $\sim$0.07\,W \\
\bottomrule
\end{tabular}
\end{table*}

In the mode comparison (Figure~\ref{fig:cs_latency_modes}), latency remains approximately constant at about 90--100\,ns over the high-voltage region, and then exhibits large excursions as voltage is reduced further. The onset of sustained latency excursions appears at approximately 0.86\,V, after which all three sweep modes exhibit frequent spikes. This aligns with the reliability transition observed in Figure~\ref{fig:cs_ber_recvsize_10g}, where BER begins rising near 0.869\,V, and with the mode dependence in Figure~\ref{fig:cs_txrx_recv_ber}, where RX-swept configurations exhibit earlier degradation than TX-only scaling.

Latency depends strongly on link speed in the stable region (Figure~\ref{fig:cs_latency_speeds}). In the nominal operating region, the measured latency is approximately 100\,ns at 10.0\,Gbps, 130\,ns at 7.5\,Gbps, 200\,ns at 5.0\,Gbps, and 410\,ns at 2.5\,Gbps. These baseline values remain approximately constant over the voltage range where each curve is stable. As voltage is reduced further, latency becomes disturbed with large excursions, starting at approximately 0.86\,V for 10.0\,Gbps, approximately 0.76\,V for 7.5\,Gbps, and approximately 0.74--0.75\,V for 5.0\,Gbps. This ordering matches the speed-dependent reliability headroom seen in Figure~\ref{fig:cs_speed_full}, where higher speeds show earlier BER onset and a narrower usable voltage range.

\subsection{Power Reduction}
\label{sec:cs_power}

We next quantify the power reduction achieved by voltage tuning on the transceiver supply rails. The goal is to separate three effects: how TX-side and RX-side power respond to rail scaling, how link speed shifts the baseline power, and how much power can be saved while remaining in the near-zero BER region, that is, before the BER onset around 0.869\,V observed in Figure~\ref{fig:cs_ber_recvsize_10g}.

\subsubsection*{\textbf{TX vs RX power trends}}

The TX-side and RX-side power traces show a clear directional dependence on which side is being voltage-scaled. When the TX-side rail is swept, TX power decreases monotonically with voltage, while keeping the TX rail fixed keeps TX power approximately constant. The same pattern holds for RX power when the RX-side rail is swept. Table~\ref{tab:cs_power_mode_summary} summarizes the observed trends and representative power levels extracted from the measured power traces.

Two practical observations follow. First, the savings from voltage tuning are largely localized to the side whose rail is being scaled, which explains why RX-side scaling has the strongest impact on reliability in Figure~\ref{fig:cs_txrx_recv_ber}. Second, the power reduction is smooth with voltage, and there is no abrupt power discontinuity at the reliability transition. Selecting an operating point is therefore fundamentally a reliability-constrained optimization rather than a power-discontinuity problem.

\subsubsection*{\textbf{Link-speed power trend}}

Link speed shifts the baseline rail power in the stable region, and all curves drop as voltage is reduced. Table~\ref{tab:cs_power_speed_summary} summarizes representative rail power values at 1.0\,V and 0.8\,V across link speeds, and we use these values to quantify both baseline differences and undervolting-induced savings.

At 1.0\,V, increasing link speed from 2.5\,Gbps to 10.0\,Gbps raises the baseline rail power by approximately 66--70\% on both TX and RX. This confirms that link speed influences not only the reliability headroom (Figure~\ref{fig:cs_speed_full}), but also the baseline operating point from which voltage tuning begins.

When undervolting from 1.0\,V to 0.8\,V, the fractional power reduction is largely similar across link speeds. On the TX side, the reduction is approximately 33--36\% across all speeds. On the RX side, the reduction is also approximately 33--35\% for most speeds, and becomes smaller at 2.5\,Gbps at approximately 25--30\%. This indicates that voltage scaling provides a comparable percentage power gain across speeds, and link speed primarily changes the baseline power level and the available voltage headroom before instability, rather than changing the efficiency of the voltage-to-power scaling itself.

This ordering matches the reliability headroom observed in Figure~\ref{fig:cs_speed_full}. Higher speeds not only show earlier BER onset, but they also operate from a higher baseline rail power. As a result, lowering the link speed offers two levers simultaneously, lower baseline rail power and a wider voltage headroom before entering the instability region.

\begin{figure*}[t]
\centering
\subfloat[BER vs power]{\includegraphics[width=0.48\linewidth]{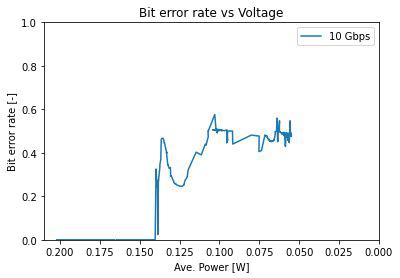}\label{fig:cs_ber_power_10g}}
\hfill
\subfloat[BER onset region (close-up)]{\includegraphics[width=0.48\linewidth]{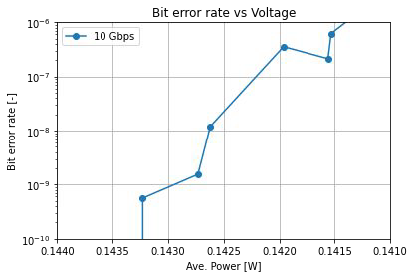}\label{fig:cs_ber_power_10g_zoom}}
\caption{BER versus rail power at 10.0\,Gbps.}
\label{fig:cs_ber_vs_power_10g}
\end{figure*}
\subsubsection*{\textbf{BER-aware power savings at 10.0\,Gbps}}

Figure~\ref{fig:cs_ber_vs_power_10g} quantifies the trade-off between link reliability and rail power at 10.0\,Gbps by plotting BER as a function of measured average rail power. The full plot shows a long near-zero-BER plateau at higher power, followed by a rapid BER increase as power is reduced. The close-up isolates this onset region and allows power savings to be quantified at specific BER levels.

First, the power saving achievable \emph{before the near-zero-BER boundary} is large. BER stays effectively zero down to approximately 0.1432\,W in Figure~\ref{fig:cs_ber_power_10g_zoom}. Taking the 1.0\,V operating point at approximately 0.20\,W as the nominal baseline in the same plot, this corresponds to approximately 28.4\% power reduction while remaining in the near-zero-BER regime.

Second, the additional power saving available in a \emph{bounded, application-tolerable BER region} is comparatively small. In the close-up, BER rises from the $10^{-10}$--$10^{-9}$ range near 0.143--0.1425\,W to approximately $10^{-7}$ near 0.1420\,W and to approximately $10^{-6}$ near 0.1415\,W. Allowing BER to increase from near-zero to approximately $10^{-6}$ therefore increases the cumulative power reduction from approximately 28.4\% to approximately 29.3\%, and the incremental gain across the onset region itself is approximately 1.2\% relative to the near-zero-BER boundary.

Third, larger power reductions are attainable only by entering the unstable regime. Below the onset region, the full BER-vs-power plot indicates that BER grows rapidly into the high-error range as power continues to drop. This behavior is consistent with the throughput collapse observed near 0.80\,V in Figure~\ref{fig:cs_ber_recvsize_10g} and with the latency-excursion regime in Figure~\ref{fig:cs_latency}, where the link becomes unstable rather than operating with bounded errors.

Overall, Figure~\ref{fig:cs_ber_vs_power_10g} shows a clear structure: most of the practical power saving comes from tuning down to the near-zero-BER boundary, approximately 28\% in this setup, while permitting bounded BER in the $10^{-7}$--$10^{-6}$ range provides only a small additional reduction before the system transitions into instability.

\section{Discussion}
\label{sec:discussion}

\subsection{Architectural Role of VolTune}
VolTune makes board-level voltage regulation a runtime-controlled architectural variable that can be invoked from within an FPGA system through a structured command interface. The PMBus control engine in Section~\ref{sec:pmbus_impl} provides a concrete, reproducible mapping from VolTune opcodes to PMBus commands and transaction sequences on KC705. The controller characterization in Section~\ref{sec:controller_eval} shows that the control plane can, in its hardware realization, be kept always-on with low overhead, while still supporting millisecond-scale voltage transitions (Figure~\ref{fig:ctrl_measure_voltage}). The representative case study in Section~\ref{sec:case_study} further shows that runtime tuning exposes an interpretable energy, reliability trade-off with measurable boundaries using BER, received data size, latency, and rail power (Figures~\ref{fig:cs_ber_recvsize_10g}--\ref{fig:cs_ber_vs_power_10g}).

\subsection{Control Mechanism and Policy Separation}
VolTune is designed as a control mechanism rather than as a fixed automatic optimizer. This is a deliberate architectural choice, because voltage-selection policy is inherently application- and platform-dependent. The present work therefore focuses on the instrumentation and command path needed to sweep voltages at runtime and to identify bounded operating regions under explicit constraints, for example near-zero BER versus bounded BER versus instability. This organization also makes the design easier to customize. Users can attach workload-specific optimization, reliability-aware control, or safety-constrained operating-point selection on top of the existing control substrate without redesigning the PMBus control engine itself. Any deployment that changes voltage during mission execution will still require a system-level policy layer that defines constraints and selects operating points based on application requirements and safety margins. This also applies to safe-voltage enforcement. VolTune provides the runtime actuation mechanism, but per-rail safety envelopes remain platform-specific and must be defined by the board configuration and the higher-level control policy used with the target device.

\subsection{Timing and Measurement Boundaries}
The achievable reaction speed is fundamentally bounded by external power-delivery dynamics and the PMBus transaction and sampling cadence. Even with a hardware control path, voltage transitions remain in the millisecond regime and measurements are derived from discrete readback samples rather than continuous analog observation (Section~\ref{sec:controller_eval}). This is sufficient for phase-level tuning and for operating-point exploration, but it is not intended for fast droop compensation or cycle-scale control. Relatedly, the settling-time definition depends on the sampling interval and stability parameters in Section~\ref{sec:ctrl_settling_method}, which should be reported and tuned consistently when comparing platforms. At the same time, the modular structure of the control plane makes such implementation choices explicit, which helps users adapt the measurement and control methodology to their own timing requirements.

\subsection{Platform Dependence and Portability}
VolTune’s platform dependence is primarily a matter of mapping rather than architecture. The PMBus transaction layer is general, but each board requires a rail map and regulator configuration, as illustrated by the KC705 lane-to-(address,\texttt{PAGE}) mapping in Table~\ref{tab:kc705_lane_map}. Porting to another PMBus-controlled FPGA board is therefore straightforward at the architectural level but still requires practical board-specific work, including identifying programmable rails, verifying safe voltage bounds, and validating measurement interpretation against the chosen regulator’s behavior. This is also an area where the design is intentionally extensible. Users can adapt the rail map, add board-specific PMBus commands, and refine the transaction layer without changing the overall architectural organization. Similarly, board-specific initialization and regulator operating-state assumptions, such as the nominal mode used before runtime tuning begins, remain part of the platform bring-up context rather than the generic VolTune runtime control abstraction.

\subsection{Case-Study Scope and Generality}
The case study also has clear scope boundaries. The transceiver rail is a strong representative target because it provides clear external metrics, but the quantitative boundaries reported here reflect a specific board, regulator configuration, link setup, and workload (Table~\ref{tab:cs_test_conditions}). The RX-dominant sensitivity observed in Figure~\ref{fig:cs_txrx_recv_ber} should be treated as an empirical result for this setup, not a universal statement about all transceiver deployments. In addition, this paper does not attempt to demonstrate multi-rail coordination or interactions between rails, and it does not model second-order effects such as rail coupling, regulator cross-effects, or thermal transients during long runs.

These scope limits do not reduce the architectural relevance of the design. Rather, they clarify the role of the present evaluation, which is to validate that VolTune can expose runtime voltage control as a practical mechanism and make application-visible trade-offs measurable on a commercial FPGA platform. The same control structure can be reused for other PMBus-controlled rails and other workloads, provided that the corresponding operating boundaries are characterized for the target setting.

\subsection{Hardware-Software Trade-offs}
The two control paths provide different but complementary integration points. The hardware control path is the better fit when the control plane must remain lightweight, deterministic, and continuously available with minimal resource cost. The software control path is useful for accessibility, rapid experimentation, and feature extension, because it simplifies changes to control logic and sequencing behavior. However, this flexibility can be expensive in FPGA memory resources and always-on static overhead (Section~\ref{sec:ctrl_overhead}). This trade-off is acceptable when BRAM is not a limiting resource and when rapid customization is needed, but it is a poor fit for BRAM-constrained accelerator designs where the control plane must remain a minimal add-on.

\subsection{Takeaway}
Overall, VolTune is best interpreted as a reusable and customizable runtime voltage-control utility block. It provides a concrete, implementable PMBus control engine and an FPGA-invocable abstraction for rail tuning, while intentionally separating low-level voltage actuation from higher-level policy decisions. This separation makes the design practical to adapt. Users who require different PMBus commands, additional telemetry, alternative settling criteria, or platform-specific control policies can extend the implementation without changing the core architectural structure. To support reproducibility and follow-on use, the implementation is released as an open-source repository containing the design artifacts used in this work \cite{voltune_repo}. Policy design, safety envelopes, and multi-rail orchestration remain explicit future work.


\section{Conclusion}
\label{sec:conclusion}

This paper presented VolTune, an open-source fine-grained runtime voltage control architecture that enables runtime tuning of FPGA supply rails through an FPGA-integrated control plane that abstracts PMBus transactions. VolTune provides both a hardware control path and a software control path, allowing designers to select between deterministic low-latency sequencing and processor-driven flexibility depending on system constraints. 

On the KC705 platform with a TI UCD9248 regulator, the hardware control path completes an end-to-end voltage transition, including regulator settling, in 2.3\,ms. The hardware-control-path overhead is small. The hardware implementation occupies 1.45\% Slice LUTs and 1.80\% BRAM tiles, and, excluding debug logic, consumes 0.015\,W of static power, corresponding to 2\% of the KC705 static power budget reported by Vivado for this subsystem.

A representative case study on GTX transceivers showed that runtime voltage tuning exposes bounded operating regions that can be identified using BER, received data size, latency, and rail power. At 10.0\,Gbps, BER remains effectively zero down to approximately 0.869\,V, then increases rapidly in the 0.869--0.864\,V region, and throughput collapses near 0.80\,V. RX-side voltage scaling dominates degradation in this setup, and the usable voltage headroom widens as link speed decreases, with BER onset shifting downward at lower rates. In terms of power, most practical savings are achieved before the near-zero-BER boundary, approximately 28.4\% in the 10.0\,Gbps setting, while permitting BER up to $10^{-6}$ yields only a small additional reduction, approximately 29.3\% total, before instability.

Future work includes integrating closed-loop voltage-selection policies on top of the existing control plane, extending the implementation to additional rails and telemetry channels, and evaluating multi-rail coordination under workload-driven tuning scenarios. 

\bibliographystyle{IEEEtran}
\bibliography{references}

\end{document}